\documentclass[aps,prl,twocolumn,groupedaddress,amsmath,amssymb,amsfonts,floatfix]{revtex4-2}

\usepackage[utf8]{inputenc}
\usepackage[english]{babel}
\usepackage[T1]{fontenc}
\usepackage{float}
\usepackage{graphicx}
\usepackage{dcolumn}
\usepackage{amsmath}
\usepackage{amssymb}
\usepackage{bm}
\usepackage[urlcolor=blue,colorlinks=true,citecolor=red,linkcolor=red,pdfstartview={FitH},bookmarks=false]{hyperref}
\usepackage{xcolor}
\usepackage{listings}
\usepackage{enumitem}
\usepackage{epstopdf}
\usepackage[normalem]{ulem}

\def\vA{{\bf A}}

\def\vBext{{\bm{B}_{\mathrm{ext}}}}

\def\Bext{{B_{\mathrm{ext}}}}
\def\Bind{{B_{\mathrm{ind}}}}

\def\BcTwo{{B_{\mathrm{c},2}}}

\def\vzhat{{\hat{\bm{z}}}}

\def\Phiext{{\Phi_{\mathrm{ext}}}}
\def\vj{{\bf j}}

\def\vF{v_{\mathrm{F}}}

\def\vpF{{\bf p}_{\mathrm{F}}}

\def\vR{{\bf R}}

\def\RCV{\mathcal{R}_{\mathrm{CV}}}
\def\pF{p_{\mathrm{F}}}

\def\NF{N_{\mathrm{F}}}

\def\thetaF{{\theta_{\mathrm{F}}}}

\def\muB{\mu_{\mathrm{B}}}
\def\kB{k_{\mathrm{B}}}

\def\Tc{T_{\mathrm{c}}}

\def\Dxtyt{{\Delta_{d_{x^2-y^2}}}}
\def\Dxy{{\Delta_{d_{xy}}}}

\def\chixtyt{{\chi_{d_{x^2-y^2}}}}
\def\chixy{{\chi_{d_{xy}}}}
\def\Dplus{{\Delta_{+}}}
\def\Dminus{{\Delta_{-}}}
\def\Dpm{{\Delta_{\pm}}}

\def\Lzorb{{\hat{L}_z^{\mathrm{orb}}}}

\newcommand{\customSection}[1]{{{\it{#1.}}---}}
\newcommand{\newtext}[1]{{\textcolor{black}{#1}}}

\allowdisplaybreaks

\begin{document}

\title{Coreless vortices as direct signature of chiral $d$-wave superconductivity}

\author{P. Holmvall}
\affiliation{Department of Physics and Astronomy,
	Uppsala University, Box 516, S-751 20, Uppsala, Sweden}
\author{A. M. Black-Schaffer}
\affiliation{Department of Physics and Astronomy,
	Uppsala University, Box 516, S-751 20, Uppsala, Sweden}

\date{\today}

\begin{abstract}
Chiral $d$-wave superconductivity has been proposed in a number of different materials, but characteristic experimental fingerprints have been largely lacking. We show that quadruply quantized coreless vortices are prone to form and offer distinctive signatures of the chiral $d$-wave state in both the local density of states and the total magnetic moment. Their dissimilarity in positive versus negative magnetic fields further leads to additional spontaneous symmetry breaking, producing clear evidence of time-reversal symmetry breaking, chiral superconductivity, and the Chern number. 

\end{abstract}

\maketitle

Exotic quantum states of matter continue to generate surprising phenomena. A prime example is chiral superconductors, where superconductivity is not only combined with non-trivial topology but also with spontaneous time-reversal symmetry breaking (TRSB) \cite{Sigrist:1991}, causing many unconventional effects \cite{Volovik:2003,Kallin:2016,Mizushima:2016,Volovik:2019,Volovik:2020}. Most outstanding is the finite Chern number set by the order parameter winding, resulting in topologically protected chiral edge modes \cite{Volovik:1997,Schnyder:2008,Hasan:2010,Qi:2011,Sauls:2011,Tanaka:2012,Graf:2013,Black-Schaffer:2014:b}.
Early focus centered on chiral $p$-wave superconductivity \cite{Kallin:2012,Kallin:2016} and its similarities to superfluidity in $^3$He \cite{Volovik:2003,Kallin:2016,Mizushima:2016,Volovik:2019,Volovik:2020}, while chiral $d$-wave superconductivity has more recently received significant attention due to proposals in a range of materials, including twisted bilayer cuprates \cite{Can:2021:a,Can:2021:b}, twisted bilayer graphene \cite{Venderbos:2018,Su:2018,Fidrysiak:2018,Xu:2018,Kennes:2018,Liu:2018,Gui:2018,Wu:2019,Fischer:2021}, ${\mathrm{Sn/Si}}(111)$ \cite{Ming:2023}, $\textrm{SrPtAs}$ \cite{Fischer:2014,Ueki:2019,Ueki:2020}, $\textrm{LaPt$_3$P}$ \cite{Biswas:2021}, ${\mathrm{Bi/Ni}}$ \cite{Gong:2017,Hosseinabadi:2019} and $\mathrm{U}{\mathrm{Ru}}_{2}{\mathrm{Si}}_{2}$ \cite{Kasahara:2007,Kasahara:2009,Shibauchi:2014,Iguchi:2021}. Furthermore, chiral $d$-wave superconductivity was recently proposed as a platform to realize topological quantum computing \cite{Mercado:2022,Margalit:2022,Huang:2023}.

Still, direct detection of both the superconducting pairing symmetry and topological invariants remain two of the most outstanding issues in physics. Consequently, undisputed discoveries of chiral superconductors have proven elusive. To make matters worse, recent studies have predicted that typical fingerprints, such as chiral edge currents and intrinsic orbital angular momentum (OAM), vanish for all pairing symmetries except $p$-wave \cite{Huang:2014,Tada:2015,Volovik:2015:b,Suzuki:2016,Wang:2018,Nie:2020,Sugiyama:2020}, further complicating measurements. Indeed, while the chiral edge modes are topologically protected, their current and OAM are not \cite{Volovik:1988,Black-Schaffer:2012,Nie:2020}. In this work, we set out to resolve this issue for chiral $d$-wave superconductors by identifying robust experimental bulk signatures in the form of distinctive vortex defects.

Vortices have been studied extensively in chiral $p$-wave superfluids \cite{Volovik:2003,Kallin:2016,Mizushima:2016,Volovik:2019,Volovik:2020}, predicting vortex defects with no analogue in conventional single-component systems. A prime example is the coreless vortex (CV) \cite{Mermin:1976,Anderson:1977,Ho:1978,Salomaa:1987,Rantanen:2023}, which is multiply quantized and non-singular with a finite superfluid order parameter everywhere. It has been sought experimentally in superfluid ${^3}\text{He}$ \cite{Hakonen:1982,Seppala:1983,Seppala:1984,Thuneberg:1986,Salomaa:1987,Parts:1995:a,Ruutu:1997,Lounasmaa:1999,Blaauwgeers:2000,Walmsley:2003}, with analogous states proposed theoretically in spin-triplet chiral $p$-wave \cite{Sauls:2009,Garaud:2012,Garaud:2015,Garaud:2016,Zhang:2016,Becerra:2016:b,Zyuzin:2017,Zha:2020,Chai:2021,Krohg:2021} and multiband \cite{Garaud:2011,Garaud:2013,Winyard:2019,Benfenati:2023} superconductors.
In comparison, however, vortices in spin-singlet chiral $d$-wave superconductors are not yet well understood, and it is not known how their higher Chern number influence vortex defects and their distinctive characteristics.

In this work we establish that CVs can easily form, without spin-triplet or multiband pairing, in chiral $d$-wave superconductors. They appear as quadruply quantized vortex defects, consisting of a closed domain wall, stabilized by eight isolated fractional vortices, and leave signatures in the local density of states (LDOS) that are easily differentiable from Abrikosov vortices. Furthermore, the chirality causes inequivalent CVs in opposite magnetic field directions, leading to further spontaneous symmetry breaking of rotational and axial symmetries in only one field direction. We show that this generates prime, smoking-gun, signatures in both LDOS and total magnetic moment that differentiates not only TRSB and chiral superconductivity, but also the orbital ordering, thus directly accessing the Chern number. These signatures are measurable using well-established experimental techniques, including scanning tunneling spectroscopy (STS) and various magnetometry setups.

\customSection{Model and method} We aim to study fundamental properties intrinsic to chiral $d$-wave superconductors, since such superconductivity has been proposed in a range of materials with widely different properties \cite{Black-Schaffer:2007,Black-Schaffer:2014:b,Can:2021:a,Can:2021:b,Venderbos:2018,Su:2018,Fidrysiak:2018,Xu:2018,Kennes:2018,Liu:2018,Gui:2018,Wu:2019,Fischer:2021,Biswas:2021,Fischer:2014,Ueki:2019,Ueki:2020,Kasahara:2007,Kasahara:2009,Shibauchi:2014,Iguchi:2021,Takada:2003,Kiesel:2013,Yamanaka:1998,Kuroki:2010,Ming:2023,Gong:2017,Hosseinabadi:2019,Mercado:2022,Margalit:2022,Huang:2023}. We thus consider a two-dimensional (2D) spin-singlet chiral $d$-wave superconductor. For simplicity, we focus here on a cylindrical Fermi surface in disc-shaped samples with radii $\mathcal{R} = 25\xi_0$, with superconducting coherence length $\xi_0 \equiv \hbar\vF/2\pi\kB\Tc$, Planck constant $\hbar$, Fermi velocity $\vF$, Boltzmann constant $\kB$, and superconducting transition temperature $\Tc$. We model clean systems with specular edges, and apply a perpendicular magnetic-flux density ${\vBext = \Bext\vzhat}$ with homogeneous flux $\Phiext = \Bext \mathcal{A}$ across the sample area $\mathcal{A}$. We assume type-II superconductivity, varying the Ginzburg-Landau constant ${\kappa \in [1,\infty)}$. \newtext{Our companion article Ref.~\cite{Holmvall:2023:a} establishes robustness of the CV and its signature in other models, e.g.~with strong anisotropy, different symmetries, Fermi surfaces, geometries, system sizes, additional vortices, and non-magnetic impurities.}

We use the well-established quasiclassical theory of superconductivity \cite{Eilenberger:1968,Larkin:1969,Serene:1983,Shelankov:1985,Eschrig:1994,Eschrig:1999,Eschrig:2000,Seja:2022,SuperConga:2023}, solving self-consistently for the order parameter $\Delta$ and vector potential $\vA$ \footnote{Our self-consistency criterion is a relative error ${<10^{-7}}$ for $\Delta$, $\vA$, $\vj$, free energy $\Omega$, and boundary condition.}, via the gap equation and Maxwell's equation,  \newtext{keeping all parameters fixed during convergence}. Apart from providing attractive pair potentials in the two $d_{x^2-y^2}$- and $d_{xy}$-wave channels \footnote{Other subdominant pair correlations compatible within group theory, e.g.~$s$-wave, are also included \cite{SuperConga:2023}.}, and providing a chiral start guess (see below), we do not constrain the superconducting state in any way. This allows the system to choose another state, e.g.~the single-component nodal $d$-wave state, but we always find the chiral state to be stable. We use a state-of-the-art implementation that runs on graphics processing units (GPUs) via the open-source framework SuperConga, already extensively used for studying vortex physics in conventional superconductors \cite{SuperConga:2023}. 

\customSection{Chiral $d$-wave superconductivity} Any 2D $d$-wave superconducting state can be described via the order parameter ${\Delta(\vpF,\vR) = \Dxtyt(\vpF,\vR) + \Dxy(\vpF,\vR)}$.
Here, $\vR$ is the center-of-mass (c.m.) coordinate and $\vpF = \pF(\cos\thetaF,\sin\thetaF)$ the Fermi momentum. Each component can be parametrized with amplitudes and phases as ${\Delta_\Gamma(\vpF,\vR) = |\Delta_\Gamma(\vR)|e^{i\chi_\Gamma(\vR)}\eta_\Gamma(\vpF)}$ with basis functions ${\eta_{d_{x^2-y^2}}(\vpF)=\sqrt{2}\cos\left(2\thetaF\right)}$ and ${\eta_{d_{xy}}(\vpF)=\sqrt{2}\sin\left(2\thetaF\right)}$. \newtext{We assume degenerate nodal components, guaranteed in three- and six-fold rotational symmetric lattices \cite{Black-Schaffer:2014:b} and thus relevant for most proposed chiral $d$-wave superconductors \cite{Venderbos:2018,Su:2018,Fidrysiak:2018,Xu:2018,Kennes:2018,Liu:2018,Gui:2018,Wu:2019,Fischer:2021,Ming:2023,Fischer:2014,Ueki:2019,Ueki:2020}. In Ref.~\cite{Holmvall:2023:a}, we consider non-degeneracy.}
A chiral $d$-wave state is characterized by a relative $\pi/2$ phase shift between these two $d$-wave components, causing TRSB \cite{Sigrist:1998}. To elucidate chirality, we re-parametrize
\begin{align}
    \label{eq:chiral:op:plus_minus}
    \Delta(\vpF,\vR) & = \Dplus(\vpF,\vR) + \Dminus(\vpF,\vR),
\end{align}
with ${\Dpm(\vpF,\vR) \equiv |\Dpm(\vR)|e^{i\chi_{\pm}(\vR)}\eta_{\pm}(\vpF)}$, where ${\eta_{\pm}(\vpF) \equiv e^{\pm i 2\thetaF}}$ is the eigenbasis of the OAM operator $\Lzorb \equiv (\hbar/i)\partial_{\thetaF}$, with eigenvalue $l^{\mathrm{orb}}_z = \pm 2\hbar$. Thus, the two components have opposite chirality and we refer to $\Dpm$ as the chiral components, while $\Dxtyt$ and $\Dxy$ are the nodal components.
 In a chiral superconductor, one of the two degenerate ground states, $\Dplus$ or $\Dminus$, becomes dominant just below $T_c$, while the other becomes subdominant. The subdominant component is fully suppressed in the bulk, but may appear at spatial inhomogeneities, such as edges, vortices, or impurities. Thus, the general form Eq.~\eqref{eq:chiral:op:plus_minus} is required when considering finite or vortex systems.
The two chiral states $\Dpm$ are fully gapped in the bulk with a non-trivial topology classified by Chern numbers $\pm2$ \cite{Volovik:1997,Schnyder:2008,Hasan:2010,Qi:2011,Sauls:2011,Tanaka:2012,Graf:2013,Black-Schaffer:2014:b}. Hence, each state hosts two chiral edge modes traversing the bulk gap, leading to finite LDOS at boundaries at all subgap energies. Chiral $d$-wave superconductors can also host domain walls, which are topological defects separating regions of opposite chirality. While they generally increase the free energy, they can be stabilized e.g.~geometrically or by disorder \cite{Garaud:2014}. Four chiral edge modes appear at a domain wall, two on each side and pairwise counter-propagating \cite{Awoga:2017}.

The best-known topological defect in superconductors is otherwise the Abrikosov vortex, associated with a $2\pi$ phase winding that locally suppresses the order parameter into a normal-state and paramagnetic core. Superconductivity recovers over the coherence length $\xi_0$ from the core, with a diamagnetic screening over the penetration depth, ${\lambda_0 \equiv \sqrt{c^2/\left(4\pi e^2 \vF^2\NF\right)}}$, with the speed of light $c$, elementary charge ${e = -|e|}$, and normal-state density of states at the Fermi level $\NF$. In a two-component order parameter, an Abrikosov vortex consists of a spatially overlapping $2\pi$ phase winding in each phase, $\chixtyt$ and $\chixy$ for chiral $d$-wave. These individual phase windings are referred to as fractional vortices, as they can carry fractional flux quantum \cite{Volovik:2000}. Spatially separating them leads to a non-singular vortex, but usually also increased energy \cite{Garaud:2013}, thus preventing separation. However, certain environments, especially a domain wall, can still favor separation. In fact, since a domain wall locally suppresses the order parameter, it typically attracts Abrikosov vortices, which can then split up into fractional vortices \cite{Sigrist:1989,Sigrist:1999}. This is the key concept for CV formation.


\begin{figure}[t!]
	\includegraphics[width=\columnwidth]{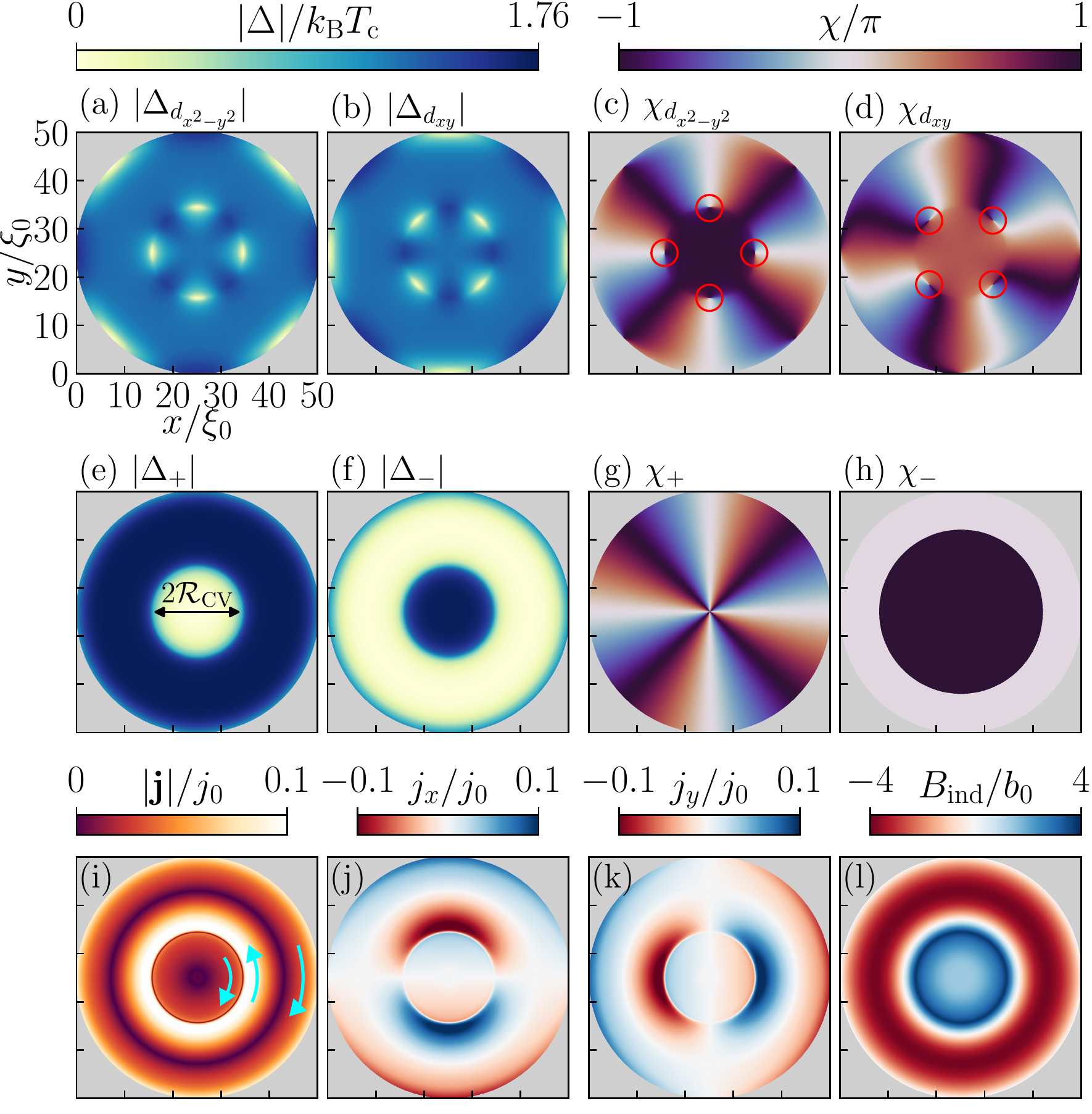}
	\caption{CV in disc with dominant chirality $\Dplus$, ${T=0.1\Tc}$, ${\Phiext=8\Phi_0}$ (${\Phi_0 \equiv hc/2|e|}$), ${\kappa=80}$, ${\mathcal{R}=25\xi_0}$. 
	(a-b) Amplitudes and (c-d) phases of the nodal components, with same in (e-h) for the chiral components. Red circles: fractional vortices.
(i) Magnitude and (j-k) components of the charge-current density ($j_0 \equiv 2\pi\kB\Tc|e|\NF\vF$). Arrows: $\vj$ direction. (l) Induced magnetic-flux density (${b_0 \equiv 10^{-5}\Phi_0/\pi\xi_0^2}$).
	}
	\label{fig:coreless_vortex}
\end{figure}

\customSection{Coreless vortex}
Figure~\ref{fig:coreless_vortex} shows a robust \footnote{CVs are extremely robust even when metastable, and appear spontaneously instead of Abrikosov vortices for many $\Phi$ and $T$. Close to $\BcTwo(T)$ the CV can become the ground state, similar to chiral $p$-wave superfluids \cite{Tokuyasu:1990:b,Garaud:2016,Krohg:2021}. A full $\Phi$-$T$ phase diagram is left as outlook.} CV computed self-consistently in a chiral $d$-wave superconductor.
Top (middle) row shows the amplitudes and phases of the nodal (chiral) order parameter components. Of central interest is the existence of four spatially separated fractional vortices in each nodal component (red circles), showing that there are no singular vortices. The fractional vortices locally suppress the corresponding nodal amplitude and lie on a circularly formed domain wall. The domain wall is clearly seen in the amplitudes of the chiral components as it separates an outer region with dominant chirality $\Dplus$ from an inner region with dominant chirality $\Dminus$, setting the CV radius ${\RCV \approx 10\xi_0}$. The dominant chiral phase $\chi_+$ winds ${-4\!\times\!2\pi}$, but the center of this winding lies in the region where already ${\Dplus \approx 0}$ (and ${\Dminus \neq 0}$). The vortex is thus non-singular and coreless. Consequently, the CV reduces the depairing caused by the external flux without paying the cost of a normal-state core, making it energetically favorable. In sharp contrast, all components are zero in Abrikosov vortex cores, see Supplemental Material (SM) for a comparison and generality of results as function of external flux. $\chi_-$ is constant besides an irrelevant $\pi$-shift, see SM.

We can understand the phase winding of the CV in Fig.~\ref{fig:coreless_vortex} by noting that a chiral state with vorticity $m$ has a phase winding ${m \times 2\pi}$ in the dominant chiral component, here with ${m=-4}$. Meanwhile, the subdominant chiral component is constrained to winding ${p = m + 2M}$ with the Chern number $M$ set by the dominant chirality, see SM for derivation. This quadruply quantized CV is the most stable CV as it corresponds to a commensurate scenario where the phase-windings from the chirality and vorticity exactly cancel in the subdominant component, ${p = -4+2\times 2 = 0}$, thus both maximizing condensation and minimizing kinetic energy. In contrast, we find that any other CV have higher energy, as $|m|\!\neq\!4$ leads to $p\neq0$, which both topologically suppresses $\Dminus$ and increases the kinetic energy, see SM. Hence, we find that a CV is very generally quadruply quantized in a chiral $d$-wave superconductor, with a total of $2|m|=8$ fractional vortices present in the nodal components, in contrast to the double quantization found in $p$-wave superfluids \cite{Sauls:2009}.

The CV generates additional interesting properties. In Figs.~\ref{fig:coreless_vortex}(i-l) we plot the resulting
charge-current density ($\vj$) and induced magnetic-flux density ($\Bind$). There is a current running in opposite directions on either side of the domain wall due to its chiral edge modes. There are also chiral edge modes at the disc edges and superposed Meissner screening currents, which generates an overall non-trivial current profile. The induced flux density shows a clear paramagnetic (blue) inner region, but in contrast to an Abrikosov vortex, the maximal paramagnetism is not at the center but at the domain wall. This magnetic ring structure offers a distinct signature for scanning magnetic probes, enhanced at lower $\kappa$.

\begin{figure}[tb!]
	\includegraphics[width=\columnwidth]{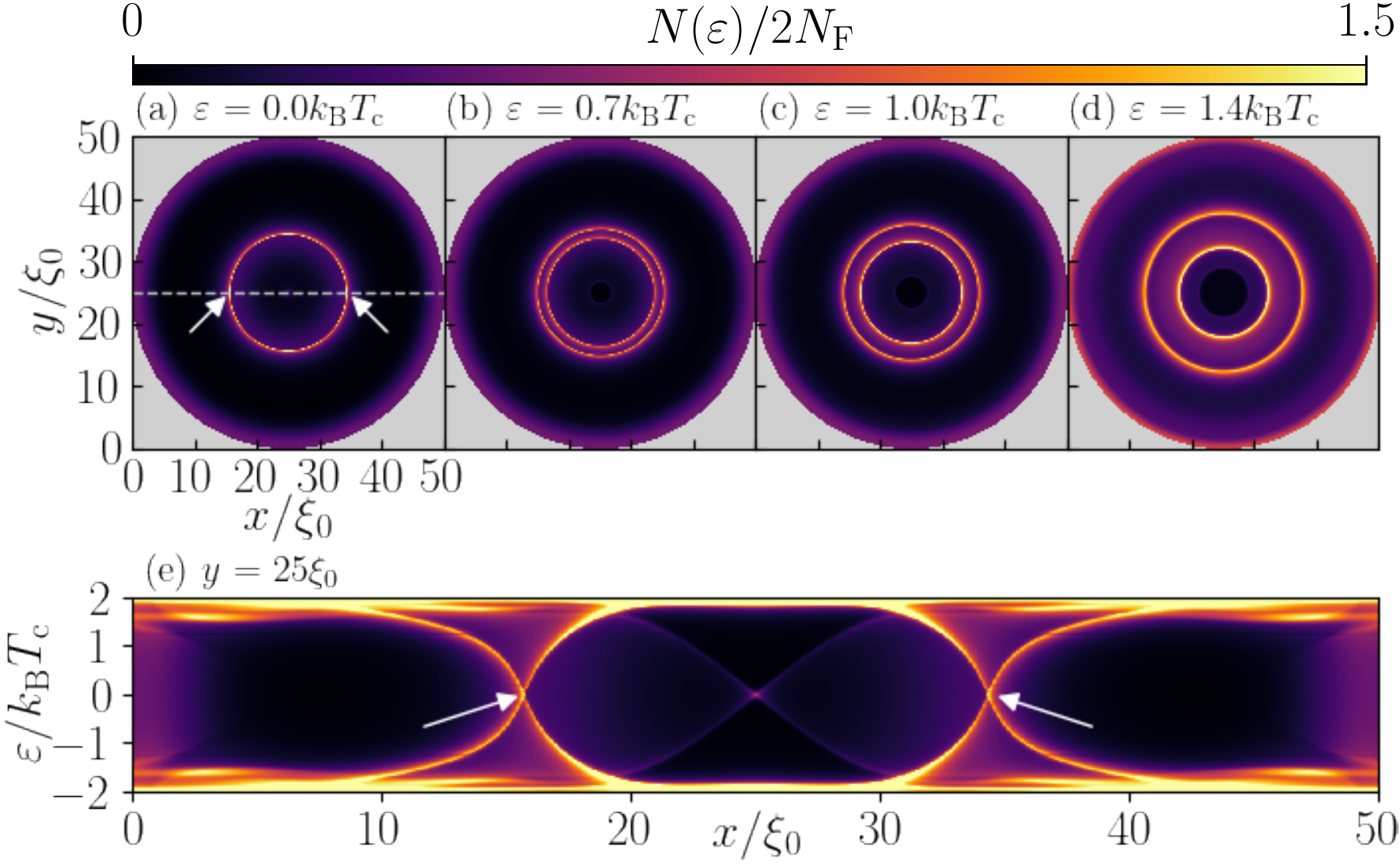}
	\caption{CV LDOS, same parameters as Fig.~\ref{fig:coreless_vortex}, and broadening $\delta\approx0.03\kB\Tc$.
        (a-d) LDOS at fixed energies $\varepsilon$. (e) LDOS along dashed line in (a). Arrows: same points in (a,e).}
	\label{fig:coreless_vortex:LDOS}
\end{figure}

We next turn our attention to the distinct LDOS signatures of a CV. Figures~\ref{fig:coreless_vortex:LDOS}(a-d) show the LDOS $N(\varepsilon)$ subgap, developing from a single ring at the domain wall at zero energy, to two concentric ring structures emanating from the domain wall at higher energies. Figure~\ref{fig:coreless_vortex:LDOS}(e) displays the LDOS along a line across the system and clearly shows how the ring separation grows with energy (or bias voltage). We attribute these ring-like states primarily to the vorticity, as a domain wall itself hosts only a small subgap DOS from the chiral edge modes, similar to the system edges seen in Fig.~\ref{fig:coreless_vortex:LDOS}(e). Notably, this stands in contrast to the point-like LDOS generated in the core of an Abrikosov vortex \cite{Caroli:1964,Rainer:1996,Berthod:2017,Kim:2021:arxiv}, see SM.

\begin{figure}[tb!]
	\includegraphics[width=\columnwidth]{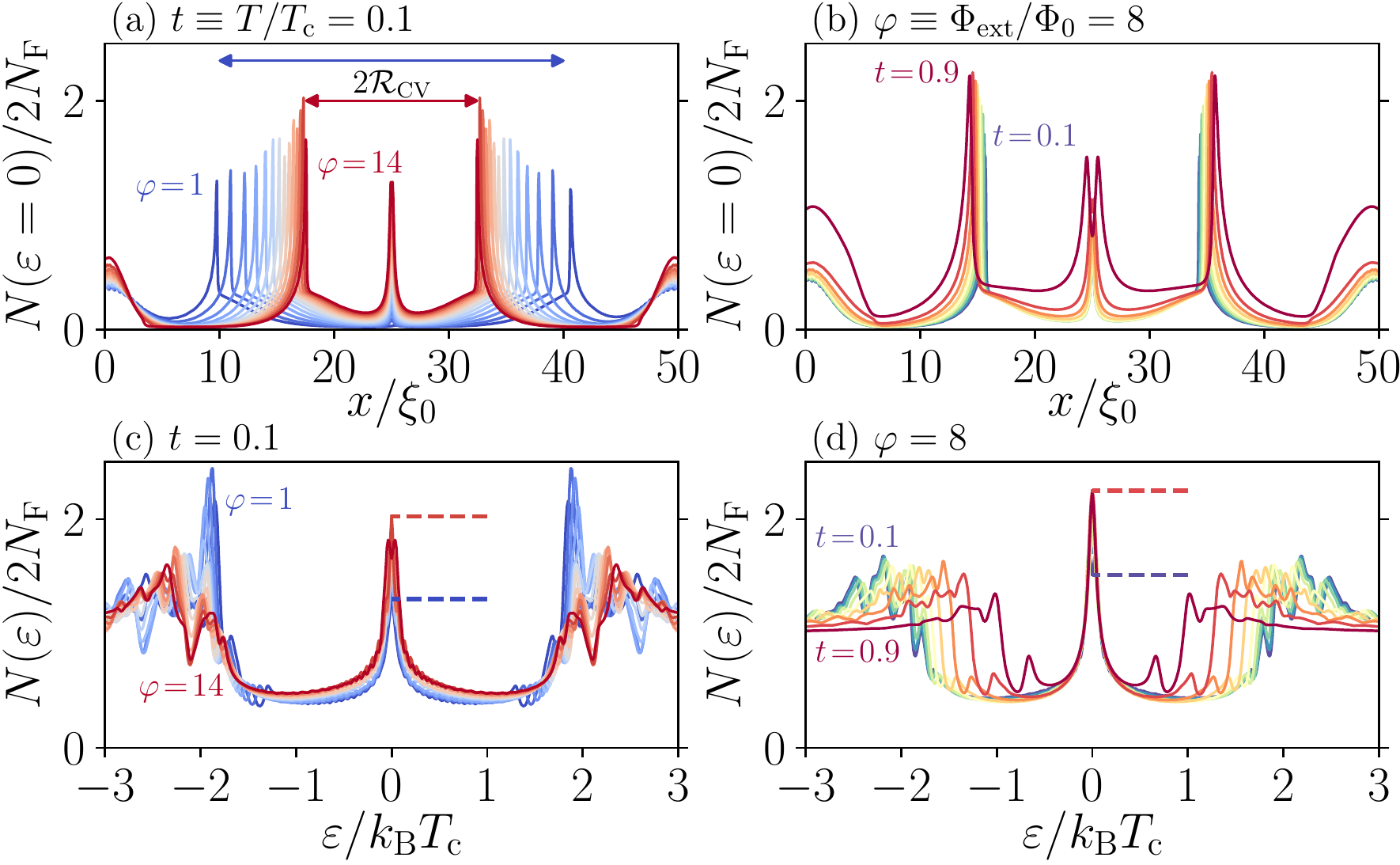}
	\caption{(a-b) Zero-energy LDOS across a CV [dashed line in Fig.~\ref{fig:coreless_vortex:LDOS}(a)]. (a) Fixed $T=0.1\Tc$, varying $\Phiext=\Phi_0$ (blue) to $14\Phi_0$ (red). (b) Fixed $\Phiext=8\Phi_0$, varying $T=0.1\Tc$ (blue) to $0.9\Tc$ (red).
	(c-d) LDOS in the domain wall of a CV [arrows in Fig.~\ref{fig:coreless_vortex:LDOS}(e)] with same parameters as (a,b). Horizontal arrows: CV size. Horizontal dashed lines: zero-energy peak height for high versus low flux (c), and temperature (d).
	}
	\label{fig:coreless_vortex:LDOS:TB}
\end{figure}

\customSection{Size and shape} 
The finite CV radius $\RCV$ is determined by competing forces \cite{Volovik:2015:a}. The repulsive interaction between the fractional vortices balances an attractive surface tension from the domain wall currents, implying that the CV size is changed by anything influencing the currents or fractional vortices. 
In Figs.~\ref{fig:coreless_vortex:LDOS:TB}(a,b) we plot the zero-energy LDOS across a CV to show how $\RCV$ is tuned directly by both the external magnetic flux and temperature. In particular, a higher flux leads to a smaller CV due to smaller separation of the fractional vortices, in analogy with denser Abrikosov vortex lattices at higher flux. Low temperature increases both the attraction and repulsion and therefore has less of an effect, but eventually leads to a small contraction. Please note, the small peak at the disc center is only present in fully symmetrical systems.
In Figs.~\ref{fig:coreless_vortex:LDOS:TB}(c,d) we extract the LDOS at the domain wall and illustrate how its zero-energy peak increases substantially with both external magnetic flux and temperature, even surpassing the coherence peaks and thus providing yet another clear signature of CVs.
Beyond temperature and flux, we find that the penetration depth $\lambda_0$ also influences the CV, by modifying the screening currents and fractional vortex separation \cite{Holmvall:2023:a}. For $\lambda_0>\mathcal{R}$ the CV radius $\RCV$ remains nearly constant (negligible screening), while for ${\lambda_0 < \mathcal{R}}$, the radius shrinks monotonically to ${\RCV\approx4\xi_0}$ at ${\lambda_0=2\xi_0}$ (strong screening). Thus, adding (non)magnetic impurities should (increase) reduce $\RCV$ \cite{Prozorov:2022}.
Beyond radius, the overall CV shape can also change. We find substantial deformation in the presence of \newtext{strong anisotropy caused by e.g.~the Fermi surface, geometry, or other vortices, see Ref.~\cite{Holmvall:2023:a}}. These results establish a strong tunability, directly reflected in the LDOS.

\begin{figure}[tb!]
	\includegraphics[width=\columnwidth]{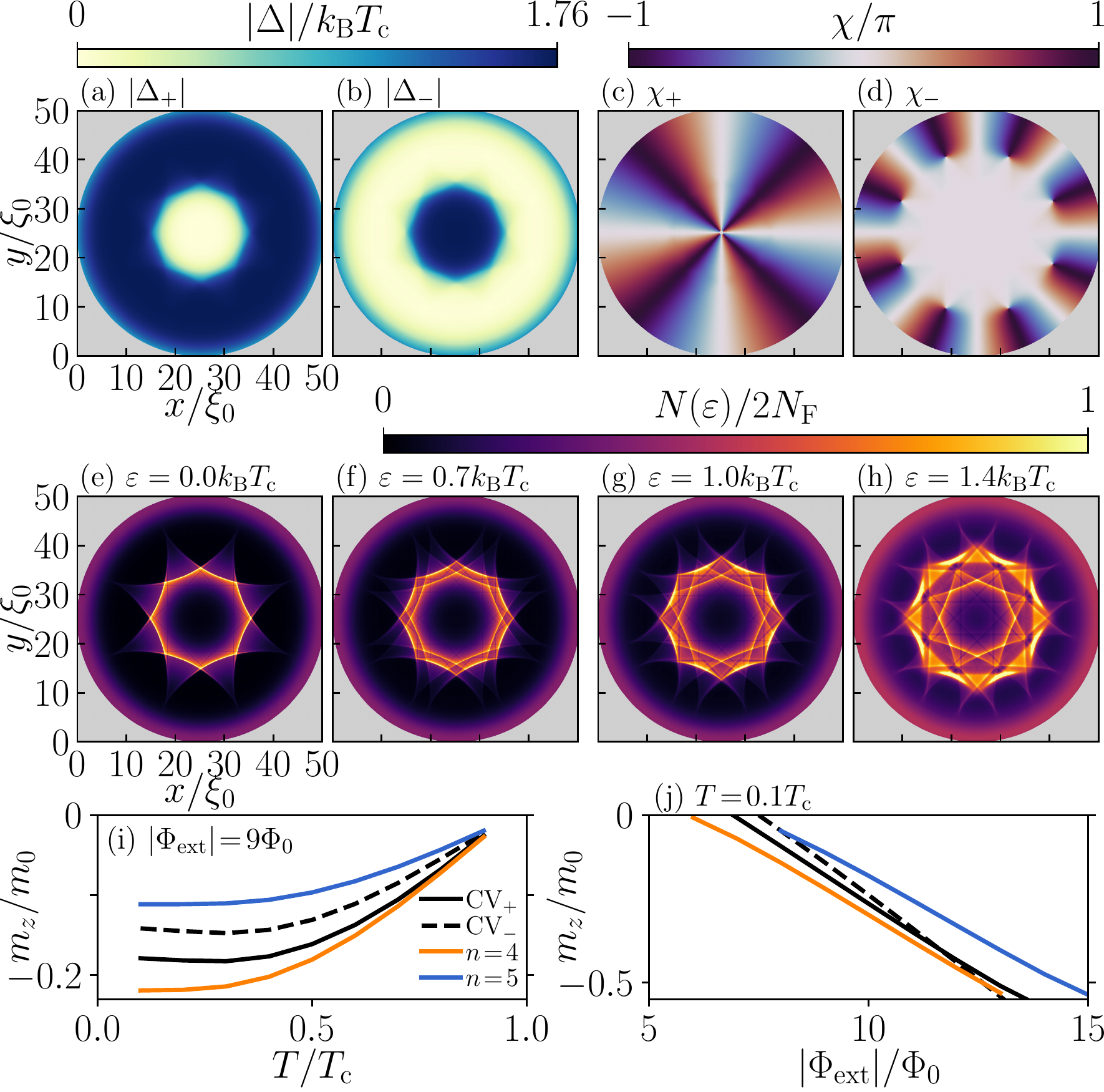}
	\caption{CV with broken axial and rotational symmetries 
        at $\Phiext=-8\Phi_0$, but otherwise the same parameters as Fig.~\ref{fig:coreless_vortex}.
        (a,b) Chiral order parameter amplitudes and (c,d) phases.
        (e-h) LDOS at fixed energies $\varepsilon$.
        (i-j) Magnetic moment $m_z$ for CV in positive flux (CV$_+$, solid lines) and $-m_z$ in negative flux (CV$_-$, dashed lines) or $n$ Abrikosov vortices ($m_0\equiv 2\muB N$ with Bohr magneton $\muB$ and particle number $N$ \cite{SuperConga:2023}).
	}
	\label{fig:coreless_vortex:ax_sym_break:collage}
\end{figure}

\customSection{Spontaneous symmetry breaking} So far, we have studied CVs in a system with dominant chirality $\Dplus$ for a positive external magnetic flux, ${\Phiext=8\Phi_0}$. Next we show that changing to negative flux leads to inequivalent CV properties, beyond simple effects of TRSB. Specifically, antiparallel (parallel) chirality and vorticity leads to canceled (enhanced) phase winding. Note that the following negative flux results are equivalent to instead changing the dominant chirality to $\Dminus$ \footnote{Results are related via the time-reversal operation $\mathcal{T}\{\vBext,\Dpm\} \to \{-\vBext,\Delta^*_\mp\}$ \cite{Garaud:2012}. Hence, results for one chirality in both field directions can be mapped to results for both chiralities in one field direction.}.
Figure~\ref{fig:coreless_vortex:ax_sym_break:collage}(a-d) shows the amplitude and phase of a CV at ${\Phiext=-8\Phi_0}$. We find again that the CV is quadruply quantized with winding $m=4$ in the dominant phase $\chi_+$. However, the subdominant $\chi_-$ has winding $p=4+2\times2=8$, since the vorticity and Chern number contributions now add, rather than cancel. This asymmetry is thus not present in a TRSB but non-chiral superconductor. Furthermore, instead of an axisymmetric winding center in $\chi_-$, there are now eight disassociated winding centers, thus spontaneously breaking axial and continuous rotational symmetries. This occurs because any axisymmetric solution would suppress $\Dminus$ at the center thus reducing superconductivity, while here the winding centers occurs where $\Dminus$ is already effectively zero. Importantly, these winding centers lead to additional phase gradients that deforms the domain wall into concave circular segments. The resulting CV in negative flux becomes a symmetry-broken and octagon-like solution with characteristic concave segments.

The broken symmetry is highly visible in the LDOS, see Fig.~\ref{fig:coreless_vortex:ax_sym_break:collage}(e-h), where the zero-energy peaks along the domain wall now takes a characteristic octagon and concave shape. More importantly, we find an even more intricate pattern at higher energies, with interweaving eight-corner concave shapes due to the eight additional winding centers, see also SM. This is thus a property due to the superposed vorticity and chirality, which by definition cannot be present in a non-chiral superconductor. Taken together, LDOS measurements on a CV in opposite field directions not only discriminate chiral from non-chiral TRSB states, but also give a direct signature of the quadruple quantization, thus measuring the $d$-wave pairing symmetry and its Chern number, while e.g.~a chiral $p$-wave state gives a square-shaped structure \cite{Sauls:2009}. \newtext{We find that these distinctive signatures survive strong broadening (see SM), non-degenerate nodal components, anisotropic Fermi surfaces, irregular systems, and non-magnetic impurities, see Ref.~\cite{Holmvall:2023:a}. The main reason for this broad generality is that chirality and vorticity are both quantized and topological.} 
Finally, in Fig.~\ref{fig:coreless_vortex:ax_sym_break:collage}(i,j) we provide complementary bulk signatures of CVs by showing the total orbital magnetic moment $m_z$ \cite{SuperConga:2023} as a function of temperature and flux, clearly different for symmetric and symmetry-broken CVs. Furthermore, both CVs produce notably different magnetic signals from Abrikosov vortices. The overall offset and slope varies between different vortex solutions, tunable by fixing temperature or flux, while measuring as a function of the other.

To summarize, we show the existence of stable quadruply quantized CVs in chiral $d$-wave superconductors. They consist of a closed domain wall with eight isolated fractional vortices, with a radius tunable by external flux, temperature, and material properties. While the CV is circular symmetric in one field direction, it spontaneously breaks rotational and axial symmetries in the other, together resulting in direct fingerprints of TRSB, chiral superconductivity, $d$-wave symmetry and thus the Chern number, in both the LDOS and magnetic moment. Beyond these smoking-gun signatures, our results also establish these superconductors as platforms for realizing fractional vortices and chiral $\mathbb{C}P^{1}$ skyrmions \cite{Babaev_Faddeev:2002, Garaud:2011,Garaud:2013,Winyard:2019,Zhang:2020,Benfenati:2023}, highly relevant in both magnetic materials and liquid crystals \cite{Nagaosa:2013,Foster:2019} as well as in particle and high-energy physics \cite{Skyrme:1962,Manton:2004,Akagi:2021,Zhang_Wang:2022:arxiv}.

We thank M.~Fogelstr{\"o}m and A.~B.~Vorontsov for valuable discussions and N.~Wall-Wennerdal, T.~L{\"o}fwander, M.~H\r{a}kansson, O.~Shevtsov, and P.~Stadler for their work on SuperConga. We acknowledge financial support from the European Research Council (ERC) under the European Union Horizon 2020 research and innovation programme (ERC-2017-StG-757553) and the Knut and Alice Wallenberg Foundation through the Wallenberg Academy Fellows program. Computations were enabled by resources provided by the Swedish National Infrastructure for Computing (SNIC) and National Academic Infrastructure for Supercomputing in Sweden (NAISS) at the computing centers NSC, C3SE, and HPC2N, partially funded by the Swedish Research Council through grant agreement No.~2018-05973.

\customSection{Note added} During the preparation of the original manuscript, a preprint \cite{Cadorim:2022:arxiv} appeared on the existence of skyrmionic chains in a twisted-bilayer model using Ginzburg-Landau theory. By using a pseudo-spin formalism \cite{Babaev_Faddeev:2002}, the CVs in our work can be seen analogously as chiral $\mathbb{C}P^{1}$ skyrmions \cite{Benfenati:2023}, which are different from skyrmionic chains, and here with different topological charge $Q=4$. Moreover, we establish the experimental accessible signatures for distinguishing time-reversal symmetry breaking, chiral superconductivity, and the Chern number.

\bibliographystyle{apsrev4-2}
\bibliography{cite.bib}

\end{document}


\title{Supplemental Material for ``Coreless vortices as direct signature of chiral $d$-wave superconductivity''}

\author{P. Holmvall}
\affiliation{Department of Physics and Astronomy,
	Uppsala University, Box 516, S-751 20, Uppsala, Sweden}
\author{A. M. Black-Schaffer}
\affiliation{Department of Physics and Astronomy,
	Uppsala University, Box 516, S-751 20, Uppsala, Sweden}

\maketitle

\normalsize
~\vspace{0.2cm} 
\renewcommand{\theequation}{S\arabic{equation}}
\renewcommand{\thefigure}{S\arabic{figure}}
\renewcommand{\figurename}{FIG.}

In this Supplemental Material (SM) we provide additional details, derivations, and results supplementing the main text of the work ``Coreless vortices as direct signature of chiral $d$-wave superconductivity", as outlined in the table of contents below. In summary, the SM illustrates the clear differences between a coreless vortex (CV) and a regular Abrikosov vortex, studies phase shifts and the phase winding constraints in the chiral order parameter components, and shows that the main results and experimental signatures are robust in the presence of strong energy broadening effects. \newtext{The robustness and tunability of the CV is further studied in the companion article, see Ref.~\cite{Holmvall:2023:a}.}

\tableofcontents

\clearpage
\section{Coreless vortex versus Abrikosov vortex}
\label{app:vortex_comparison}
In this SM section we report on the properties of Abrikosov vortices to provide a direct comparison to the CVs studied in the main text. In order to ease comparisons and highlight the many differences, we provide similar plots of order parameters and LDOS as Figs.~1 and 2 in the main text. We also report detailed line-cuts of order parameter, currents, and induced magnetic flux-density through a CV, Abrikosov vortex, and a system with no vorticity.

\begin{figure*}[b!]
    \centering
    \begin{minipage}[t]{0.49\textwidth}
	\includegraphics[width=\columnwidth]{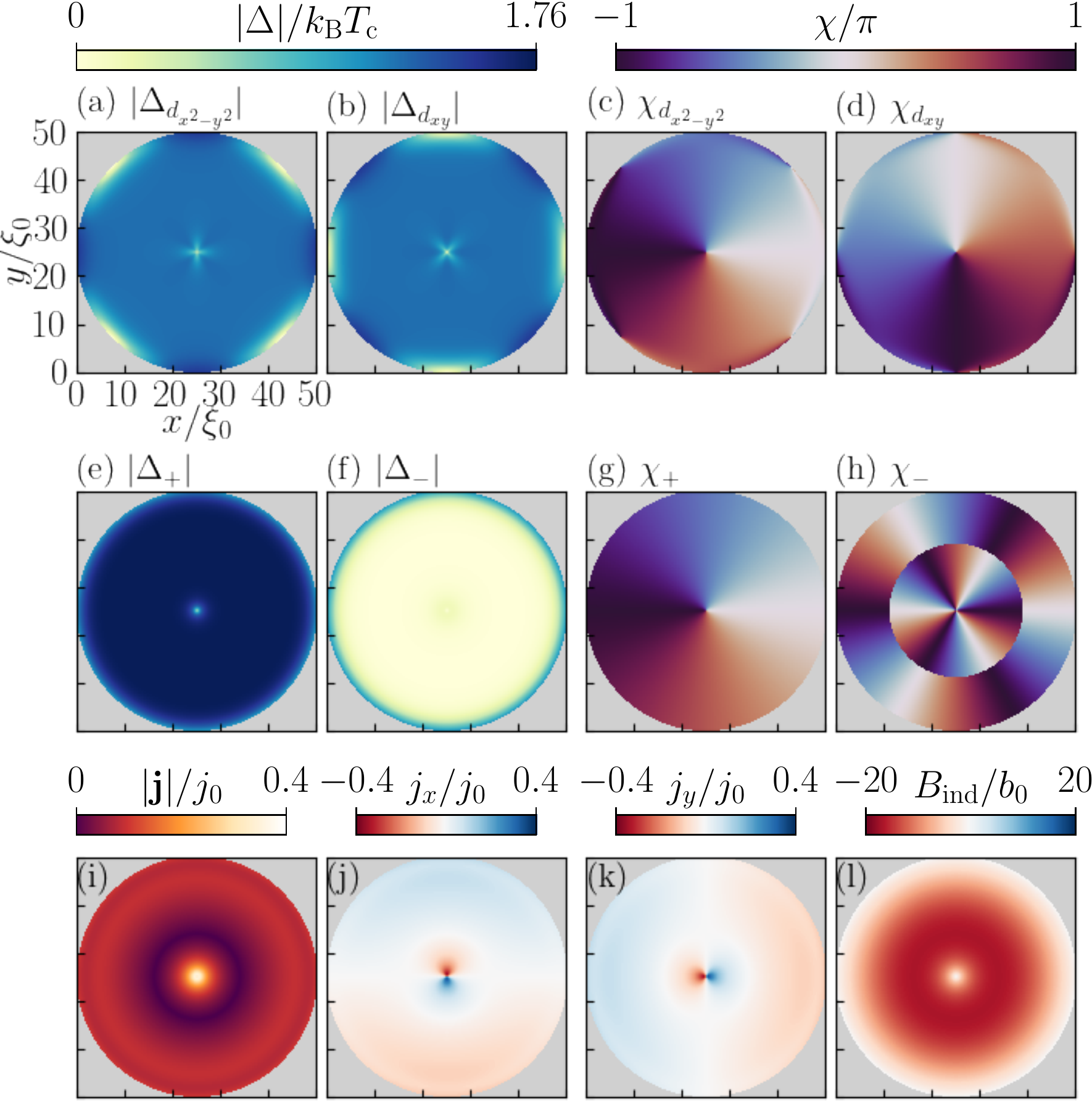}
	\caption{Same as Fig.~1 in main text but for an Abrikosov vortex in system with dominant chirality $\Dplus$ at $T=0.1\Tc$, $\Phiext=8\Phi_0$, $\kappa=80$, $\mathcal{R}=25\xi_0$.
	(a-b) Amplitudes and (c-d) phases of the nodal components, with the same in (e-h) but for the chiral components.
(i) Magnitude and (j-k) components of the charge-current density. (l) Induced magnetic-flux density.
	}
	\label{fig:abrikosov_vortex}
    \end{minipage}\hfill
    \begin{minipage}[t]{0.49\textwidth}
	\includegraphics[width=\columnwidth]{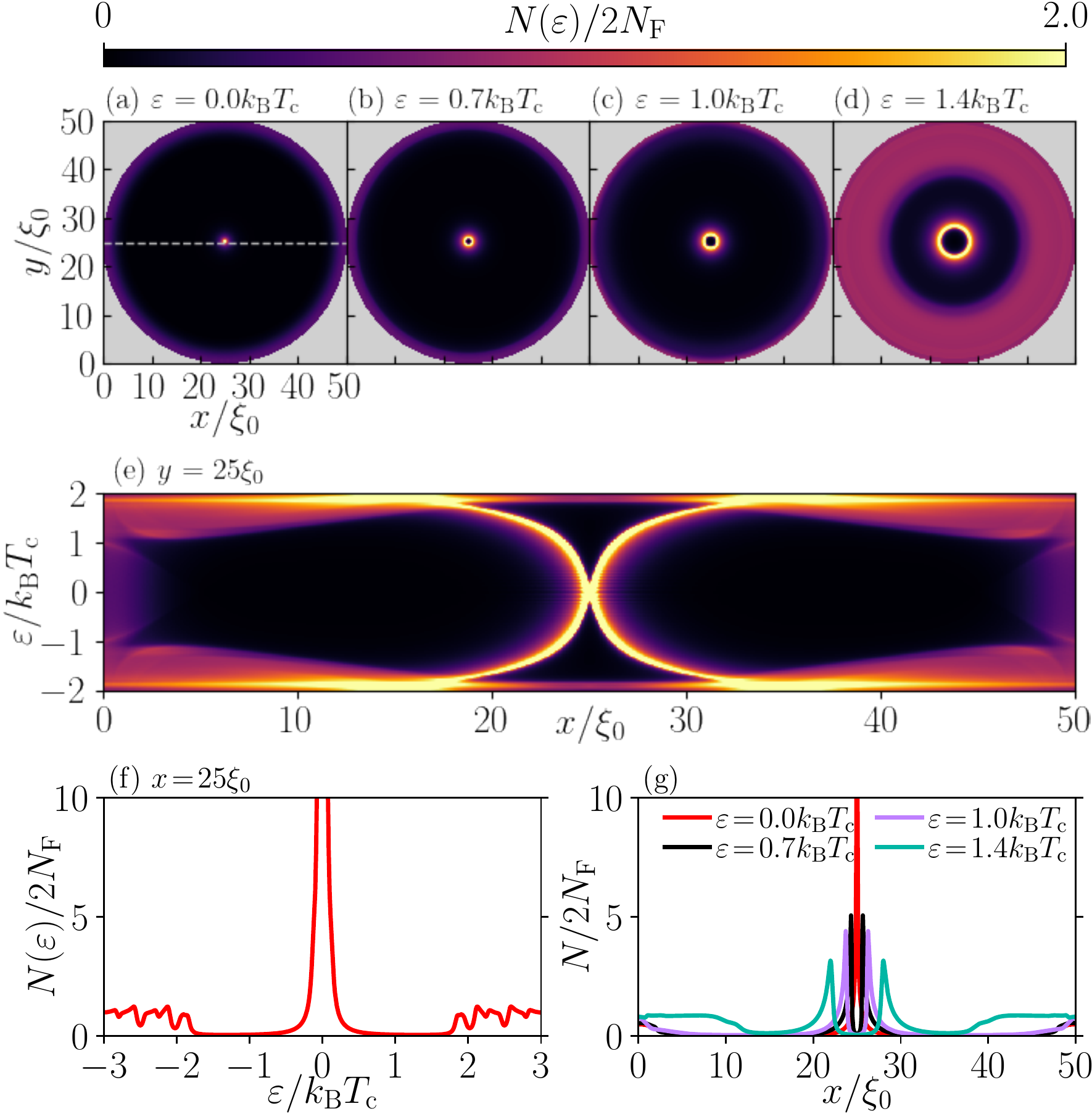}
	\caption{Same as Fig.~2 in main text but LDOS for an Abrikosov vortex and with additional line cuts. Same parameters as Fig.~\ref{fig:abrikosov_vortex}.
        (a-d) LDOS versus spatial coordinates $x$ and $y$ at fixed energies $\varepsilon$. (e) LDOS across dashed line in (a). LDOS in the vortex core (f) and across dashed line in (a) at fixed $\varepsilon$ (g). Energy broadening set to $\delta \approx 0.03 k_B T_c$.
	}
	\label{fig:abrikosov_vortex:LDOS}
    \end{minipage}
\end{figure*}

\begin{figure*}[b!]
	\includegraphics[width=\textwidth]{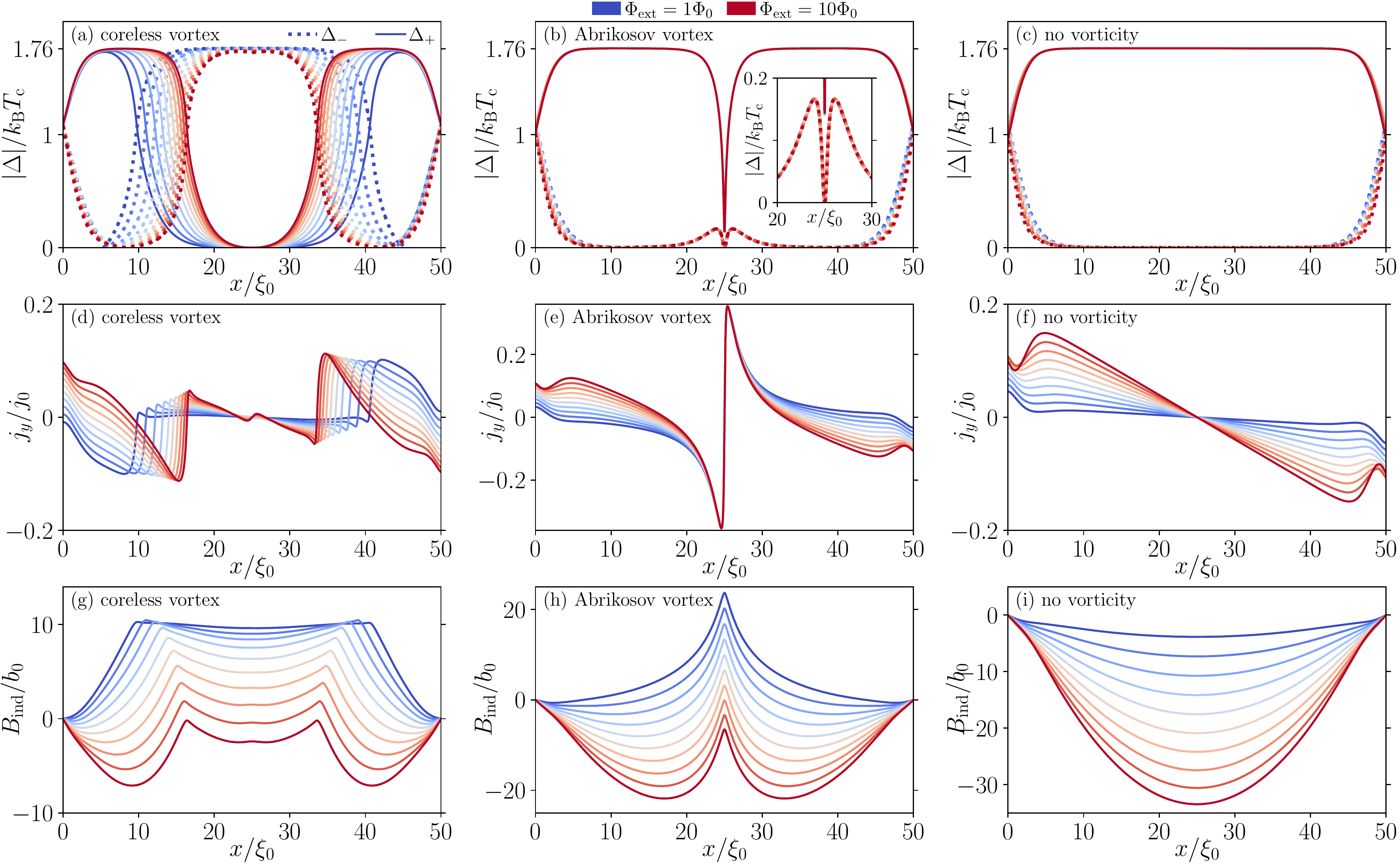}
	\caption{Line-cuts through CV (first column), Abrikosov vortex (middle column), and systems without vorticity (last column) with first row displaying the order parameter amplitude, second row the surface-parallel component of the charge-current density, and third row the induced magnetic-flux density. Line colors indicate different external flux from $\Phiext = \Phi_0$ (blue) to $\Phiext = 10\Phi_0$ (red) in steps of $\Phi_0$. All other parameters the same as Figs.~1 in the main text and \ref{fig:abrikosov_vortex}.
	}
	\label{fig:coreless_vortex:cuts}
\end{figure*}

Figure~\ref{fig:abrikosov_vortex} shows the spatial profile of an Abrikosov vortex, with the same layout as the CV in Fig.~1 in the main text. Figures~\ref{fig:abrikosov_vortex}(a-d) show that the suppression and phase winding of the nodal order parameter components occur at the same coordinate for an Abrikosov vortex, in contrast to the CV, while (e-f) show that both chiral components are exactly zero in the core, again different from the CV (although in a small region of radius $\sim\xi_0$ around the singular Abrikosov vortex core, the subdominant chirality is locally induced, i.e.~$\Dminus\neq0$, more clearly seen in Fig.~\ref{fig:coreless_vortex:cuts}(b)). The Abrikosov vortex has a phase winding ${m=-1}$ in the dominant chiral phase ($\chi_+$), which leads to a constrained subdominant chiral phase ($\chi_-$) with winding ${p=m+4=3}$. The additional ring visible in the structure in $\chi_-$ corresponds to a $\pi$ phase shift, just like in the CV (see later SM section for an analysis). Figures~\ref{fig:abrikosov_vortex}(i-l) show that the Abrikosov vortex core hosts paramagnetic currents and paramagnetic induced flux density, the latter appearing as white instead of blue colors due to the relatively large external flux ($\Phiext=8\Phi_0$) and penetration depth ($\kappa=80$). The core is surrounded by opposite current circulation with diamagnetic screening (red).

In Fig.~\ref{fig:abrikosov_vortex:LDOS} we plot the LDOS of an Abrikosov vortex with the same layout as Fig.~2 in the main text for a CV, and with additional line cuts in (f,g).
The normal core of an Abrikosov vortex is well-known to host subgap Caroli-de-Gennes-Matricon states \cite{Caroli:1964,Rainer:1996,Berthod:2017,Kim:2021:arxiv}. In a topological superconductor, such as the chiral $d$-wave state, there are additional zero-energy states in the vortex core center \cite{Alicea:2012}. The zero-energy states are clearly visible in Fig.~\ref{fig:abrikosov_vortex:LDOS}(a) and stand in sharp contrast to the large ring structure of the CV. As the energy (bias voltage) increases in (b-e), the zero-energy states in the Abrikosov vortex develop into a small circular peak of Caroli-de-Gennes-Matricon subgap states, which carry the paramagnetic currents \cite{Rainer:1996,Fogelstrom:2011,Holmvall:2017}.
For energies below $\varepsilon<\kB\Tc$, the circular peak structure has a maximal radius of $\sim\xi_0$ and is therefore an order of magnitude smaller than the ring structures found in the LDOS of the CV. Furthermore, comparing the LDOS across the disc diameter in Fig.~\ref{fig:abrikosov_vortex:LDOS}(e), we find two branches crossing the gap for the Abrikosov vortex, while there are four for the CV. Not only the number of branches are different, Figs.~\ref{fig:abrikosov_vortex:LDOS}(f,g) highlight that the magnitude is also about two orders of magnitude larger in the Abrikosov vortex as compared to in the CV in Fig.~3 in the main text. A major contributing factor to this is the stronger pair-breaking in the fully normal core of the Abrikosov vortex. Taken together, these contrasting LDOS results provide strong, experimentally accessible, signatures to distinguish between CVs and Abrikosov vortices.

Finally, in Fig.~\ref{fig:coreless_vortex:cuts} we provide additional line cuts through the center of a CV (left column), an Abrikosov vortex (middle column), and a system with no vorticity (right column). 
The top row shows the chiral order parameter amplitudes, the middle row the surface-parallel component of the charge-current density, and the bottom row the induced magnetic-flux density. Different lines correspond to different values of the external flux from $\Phiext=1$ (blue) to $\Phiext=10$ (red). Most striking is how strongly the order parameter profile (and therefore radius) of the CV varies with external flux, compared to the other two systems, where there is no notable variation.
Consequently, the current density varies only in magnitude as a function of flux in the systems with the Abrikosov vortex or no vortex, while both the magnitude and spatial extent varies for the CV. For the induced magnetic flux density, the CV has a clear paramagnetic ring structure for all lower flux densities, while the Abrikosov vortex only has a narrow paramagnetic point and the system with no vorticity is always diamagnetic, as expected. We note that the penetration depth is here relatively large compared to the disc radius, leading to weak screening and small induced flux. A smaller penetration depth leads to a much stronger induced flux density (especially when ${\lambda_0 < \mathcal{R}}$ and ${\lambda_0} \sim 10\xi_0$), with more pronounced ring-like versus point-like structures, and therefore an even larger difference between the systems.

In summary, the above results show how the CV and Abrikosov vortex have remarkably different properties that leads to clearly measurable differences in local-probe experiments, such as scanning tunneling spectroscopy (STS) \cite{Hess:1989,Renner:1991,Maggio-Aprile:1995,Yazdani:1997,Pan:2000a,Pan:2000b,Hoffman:2002,Guillamon:2008,Roditchev:2015,Berthod:2017} and various magnetometry setups \cite{Geim:1997,Tsuei:2000,Morelle:2004,Kirtley:2005,Khotkevych:2008,Kokubo:2010,Bert:2011,Vasyukov:2013,Curran:2014,Ge:2017,Persky:2022}.

\clearpage
\section{Phase shift in the subdominant chiral component}
\label{app:pi_jump}
In this SM section we analyze the $\pi$-valued phase shift that appears in the phase of the subdominant chiral component, in $\chi_-$ for both the CV in Fig.~1(h) and  the Abrikosov vortex in Fig.~\ref{fig:abrikosov_vortex}(h). 

\begin{figure*}[b!]
    \centering
    \begin{minipage}[t]{0.49\textwidth}
	\includegraphics[width=\columnwidth]{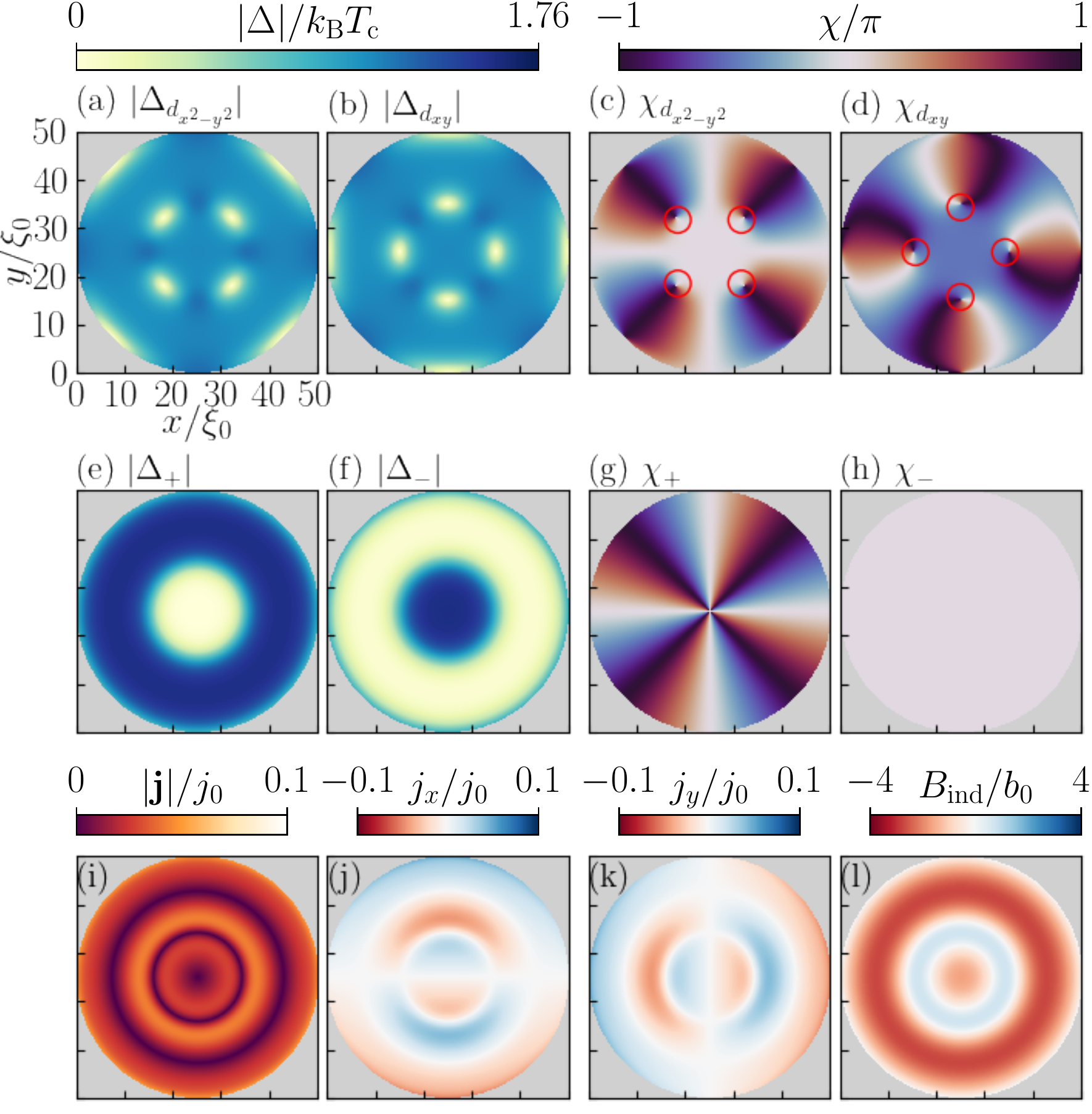}
	\caption{Same as Fig.~1 in main text but for a CV without a $\pi$-shift in the subdominant chirality at same parameters as Fig.~1 except $T=0.6\Tc$ (needed to provide a stable solution).
	}
	\label{fig:coreless_vortex:no_pi}
    \end{minipage}\hfill
    \begin{minipage}[t]{0.49\textwidth}
	\includegraphics[width=\columnwidth]{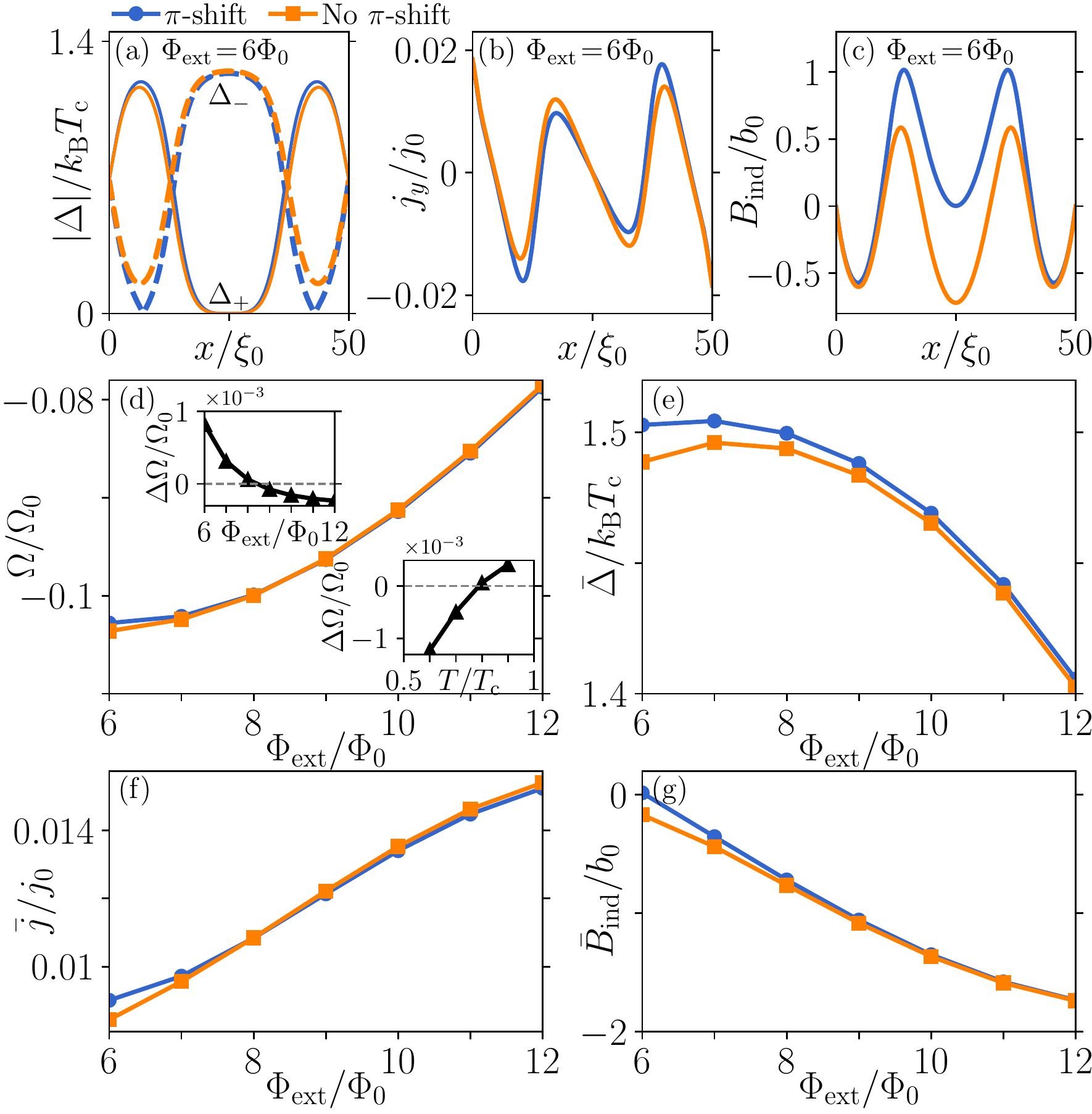}
	\caption{Comparison of CVs at $T=0.8\Tc$, with a $\pi$-shift (blue) and without it (orange) in the subdominant chiral phase $\chi_-$. Line-cuts through the CV of the order parameter amplitudes $|\Delta_+|$ (solid) and $|\Dminus|$ (dashed) (a), charge-currenty density (b), and induced magnetic-flux density (c).
 (d) Total free energy $\Omega$ ($\Omega_0 \equiv \mathcal{A}\NF(\kB\Tc)^2$). Insets: free-energy difference $\Delta\Omega$ with and without $\pi$-shift. Area-averaged order parameter amplitude (e), charge-current density (f), and induced magnetic-flux density (g). }
	\label{fig:coreless_vortex:no_pi:cuts}
    \end{minipage}
\end{figure*}

In Fig.~\ref{fig:coreless_vortex:no_pi} we plot a CV without the $\pi$-shift, with the same parameters and layout as in Fig.~1 in the main text, but using a higher temperature ($T=0.6\Tc$ instead of $T=0.1\Tc$) necessary to stabilize this solution. Figures~\ref{fig:coreless_vortex:no_pi}(a-d) show that a lack of a $\pi$-shift leads to a rotation of the nodal order parameters, such that the fractional vortices are aligned with the order parameter lobes (nodes) when the $\pi$-shift is present (absent). For larger systems, this rotation is necessarily trivial in the rotationally symmetric CV. However, in smaller systems this might lead to a finite hybridization between the fractional vortices and a system boundary-induced suppression of the nodal components, i.e.~see the shorter distance between regions of suppressed amplitudes (white) in Figs.~\ref{fig:coreless_vortex:no_pi}(a-b) compared to in Figs.~1(a-b). In (e-l), we only notice an overall quantitative difference due to the different temperatures.

Figure~\ref{fig:coreless_vortex:no_pi:cuts} provides a more quantitative comparison, showing results with the $\pi$-shift (blue) and without it (orange). First in (a-c) via spatial line cuts at $T=0.8\Tc$ and $\Phiext=6\Phi_0$ and then in (d-g) as area-averaged quantities (denoted by the ``bar'' symbol, e.g.~ $\bar{\Delta}$) as a function of external flux at $T=0.8\Tc$, including the free energy in (d). We begin by pointing out that the $\pi$-shift in $\chi_-$ takes place in the region where the order parameter amplitude $|\Dminus|$ is exponentially suppressed. Interestingly, in small systems where the typical suppression length is larger than the distance between the CV and the system edge, the $\pi$-shift acts as an additional topological defect that enforces an exact suppression in $|\Dminus|$ between the domain wall and the system edge, which would have naturally been present in a larger system. Still, the $\pi$-shift can have a small effect also in larger systems, since it induces a non-trivial Josephson coupling between two regions where ${\Dminus\neq0}$; the center and edge of the disc. This influences the charge-current density (b), which in turn influences the induced magnetic flux-density and hence the screening in (c). 
In (d) we report the total free energy $\Omega$, computed with respect to the normal state, via the Eilenberger free-energy functional \cite{SuperConga:2023,Eilenberger:1968,Virtanen:2020} in order to discern the most stable configuration. Insets show the free-energy difference between having and not having the $\pi$-shift, as a function of flux at fixed temperature $T=0.8\Tc$ (upper inset) and as a function of temperature at fixed flux $\Phiext=8\Phi_0$ (lower inset). Together, these results illustrate that the $\pi$-shift solution is energetically favorable at all lower temperatures and higher flux. For this large parameter regime, we find the $\pi$-shift to be very robust and appearing spontaneously in our simulations, while a CV without the $\pi$-shift is much harder to stabilize. We trace this flux-temperature dependence to both the order parameter and current. Around $(\Phiext,T)=(8\Phi_0,0.8\Tc)$, where the free-energy difference is smallest, the $\pi$-shift has a negligible impact on both of these quantities, as indicated in (e-g). Far from this point, in particular at low temperature or flux, a somewhat larger deviation starts appearing in the total order parameter in (e) and the current density in (f), which leads to a modified induced magnetic-flux density (g). The latter influences also the screening behavior. In turn, these terms influence the condensation energy, kinetic energy, and magnetic-energy density. The most favorable configuration is thus determined by the balancing of these contributions. Further analysis of this balance is quite technical and not important for the main message or results in this work, and we therefore leave it as an outlook.

To summarize, the above results show that a $\pi$-shift in the subdominant order parameter is favorable at all low temperatures and high fluxes. However, it has only a minor effect on the overall properties of CV, and as such are not important for the overall analysis of the CV properties.

\clearpage
\section{Phase winding constraint and vorticity in chiral superfluids}
\label{app:phase_winding_and_vorticity}
In this SM section we study the phase winding of a chiral superfluid, in order to better understand the coreless vortex in the main text. We derive the constraint between the phases of the dominant and subdominant chiral components for a chiral $d$-wave order parameter. We show that vorticity leads to a center-of-mass (c.m.) angular momentum and a phase winding of the dominant component, while chirality is also associated with an intrinsic orbital angular momentum (OAM) causing an additional phase winding in the subdominant component. This highlights that phase winding (in the subdominant component) does not always imply vorticity in a chiral superfluid. The following derivation is a generalization of the calculations for chiral $p$-wave in Ref.~\cite{Sauls:2009}, which contains a more in-depth analysis.

We consider a chiral superfluid of ``chiral order'' $|M|$, corresponding to the Chern number, and write the components of the order parameter in the eigenbasis of the OAM operator, as in the main text, $\Dpm(\vpF,\vR) = |\Dpm(\vR)|e^{i\chi_\pm(\vR)}e^{\pm i|M| \thetaF}$.
Even (odd) $M$ corresponds to spin singlet (triplet) superconducting states and $|M|=1,2,3,\ldots$  generate chiral $p,d,f,\ldots$-wave symmetry. This state is an eigenstate of the OAM operator $\Lzorb \equiv (\hbar/i)\partial_{\thetaF}$, $\Lzorb \Dpm(\vR,\vpF) = l_{\textrm{orb}} \Dpm(\vR,\vpF)$ with eigenvalues $l_{\textrm{orb}} = \pm \hbar |M|$.
This corresponds to a condensate that carries a net OAM along the $\pm \hat{z}$-axis \cite{Sauls:1994}. In contrast, a non-chiral state, like the nematic $d$-wave state or even a TRSB $d+is$-wave state, is not an eigenstate of $\Lzorb$. Consider next the same superfluid in the presence of flux vortices due to an external applied magnetic field. Far from any vortex core or other significant inhomogeneity, the subdominant chirality is fully suppressed and the order parameter approaches the asymptotic bulk value $\Delta(\vpF,\vR) \approx \Delta_\pm(\vpF,\vR)$, but now with a topologically non-trivial phase winding $\chi_\pm(\vR) = m \phi$. Here, $\phi$ is the polar angle of the c.m. coordinate $\vR = |\vR|(\cos\phi,\sin\phi)$, and $m$ is the integer vorticity with its sign set by the direction of the external magnetic field. Specifically, $m<0$ for vortices ($m>0$ for antivortices) in positive external flux, and vice versa in negative flux. Since each order parameter component is a single-valued quantity, the phase has to wind $2\pi \times m$ along a contour encircling all vortices. Hence, $m$ can be referred to as the \textit{global} winding number, carried by the dominant chirality. As a result, the Cooper pairs also carry a c.m. angular momentum, obtained from the operator $\Lzcm \equiv (\hbar/i)\partial_\phi$, with eigenstate $\Lzcm\Dpm(\vR,\vpF) = l_{\textrm{c.m.}}\Dpm(\vR,\vpF)$ and eigenvalue $l_{\textrm{c.m.}} = \hbar m$. 
Asymptotically far from the vortices, the total angular momentum of each Cooper pair is obtained via $\hat{L} = \Lzcm + \Lzorb$ \cite{Nie:2020} and thus becomes $l = l_{\textrm{c.m.}} + l_{\textrm{orb}} = \hbar \left( m \pm |M| \right)$, for the two chiralities $\Dpm$. Close to the vortex cores, the subdominant component $\Dmp(\vR,\vpF)$ is locally induced \cite{Heeb:1990,Ichioka:2002}, with a total phase winding $\chi_\mp(\vR) \to p\phi$, where $p$ can thus be referred to as the \textit{local} winding number. Acting with $\Lz$ on the total order parameter $\Delta(\vR,\vpF) = \Dpm(\vR,\vpF) + \Dmp(\vR,\vpF)$ yields $\Lz \Delta(\vR,\vpF) = \hbar \left(m \pm |M|\right)\Dpm(\vR,\vpF) +\hbar\left(p \mp |M|\right)\Dmp(\vR,\vpF)$. This state is still fully chiral and thus an eigenstate of $\Lz$ \cite{Sauls:2009}, resulting in the condition $l = \hbar(m \pm |M|) = \hbar(p \mp |M|)$. Hence, we get a constraint between the global (dominant) phase $m$ and the local (subdominant) phase $p$
\begin{align}
    \label{eq:chiral:phase_constraint}
    p = m \pm 2|M|.
\end{align}
 Even in the absence of vorticity ($m=0$), the subdominant component thus has a finite phase winding $p=\pm2|M|$ due to the intrinsic OAM. Hence, for a chiral $d$-wave state with $|M|=2$, we find that $p = m \pm 4$, where the sign is set by the dominant chirality in the system ($\Dpm$). This expression can be compared to the result derived for a chiral $p$-wave superconductor: $p = m \pm 2$ \cite{Sauls:2009}. We further note that the quadruply quantized and symmetric CV found as the most stable solution and discussed the main text is a commensurate state where the contribution from the vorticity (i.e.~c.m.~angular momentum) and Chern number (i.e.~OAM) exactly cancel: ${p=-4+2\times 2 = 0}$. The reason for this being the most stable solution is that any other vorticity causes a finite phase winding $p$, which is energetically less favorable, as it both suppress superconductivity within the core and increase the kinetic energy as soon as the vorticity is increased. The opposite field direction instead leads to large winding $p=4+2\times2=8$. Still, this state partially avoids the energy penalty of suppressed superconductivity by placing the additional phase winding outside the coreless vortex where the subdominant component is already zero, but which results in  spontaneously broken axial symmetry. These additional, off-centered, winding centra drive additional superflow, which distorts the CV domain wall. This asymmetry and associated distortion are therefore directly related to the Chern number and orbital angular momentum of the chiral state, consequently, also responsible for producing the clear experimental signature shown in the main text. Finally, we note that there might be additional contributions due to higher-order harmonics when rotational symmetry is broken, e.g.~by non-circular Fermi surfaces. For the point group $D_{4h}$, this amounts to an extra term $+4n$ in Eq.~(\ref{eq:chiral:phase_constraint}), with integer $n$ \cite{Yip:1993}. However, such terms have been found to often lead to a significant increase in the free energy \cite{Sauls:2009}, and are thus not favored. Regardless, we demonstrate that our main results are robust in the presence of such higher-order harmonics, see companion article \cite{Holmvall:2023:a}. We thus conclude that the clear LDOS signature observed in the main text is directly connected to the interplay between the two topological and quantized properties, namely chirality and vorticity.

\clearpage
\section{Additional plots for symmetry-broken coreless vortices}
\label{app:cv_asb:additional_results}
\begin{figure*}[b!]
    \centering
    \begin{minipage}[t]{0.49\textwidth}
	\includegraphics[width=\columnwidth]{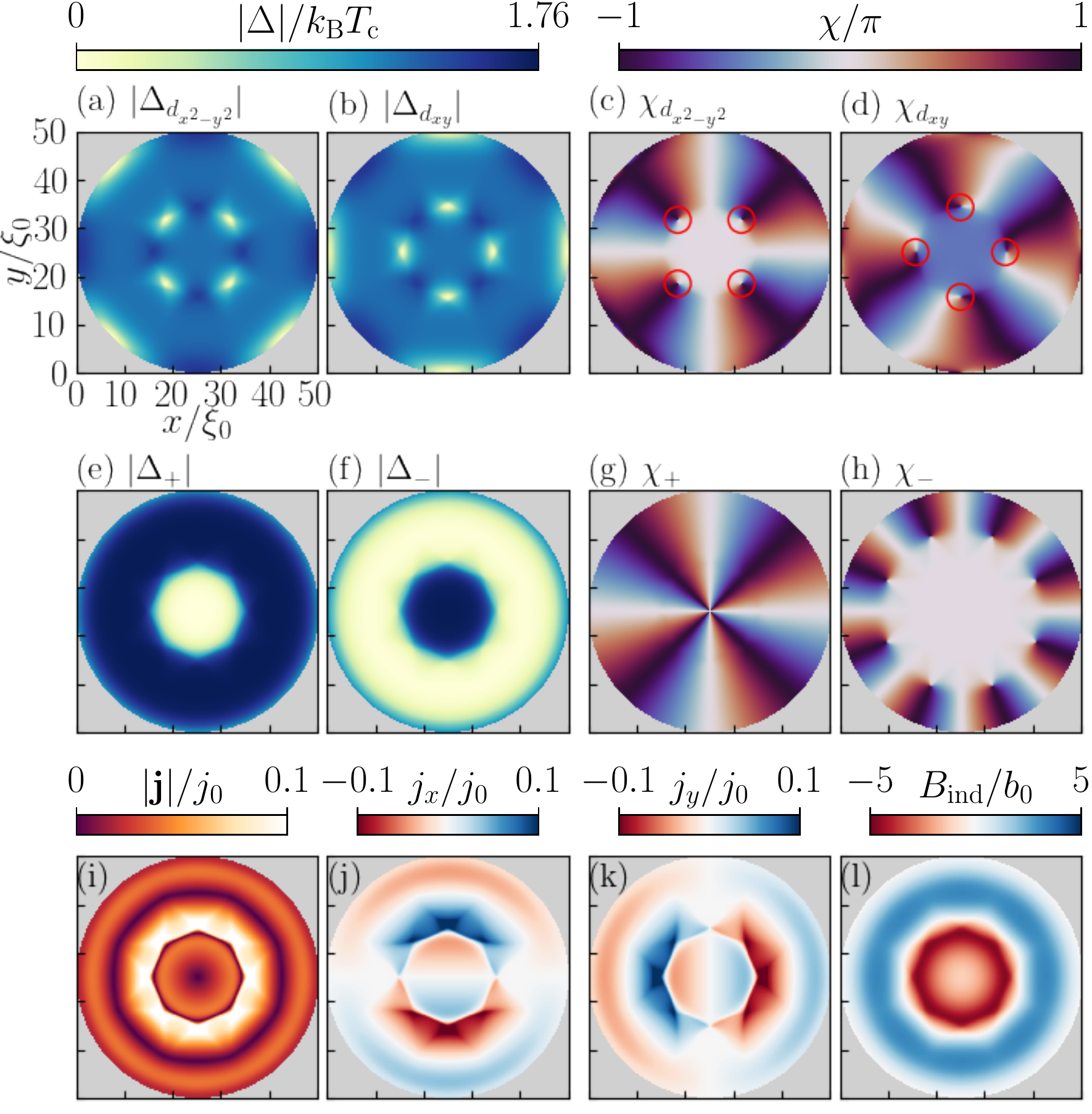}
	\caption{Same as Fig.~1 in main text but for a CV with broken axial and rotational symmetries at $\Phiext=-8\Phi_0$ but otherwise the same parameters and layout.
	}
	\label{fig:coreless_vortex:ax_sym:break}
    \end{minipage}\hfill
    \begin{minipage}[t]{0.49\textwidth}
	\includegraphics[width=\columnwidth]{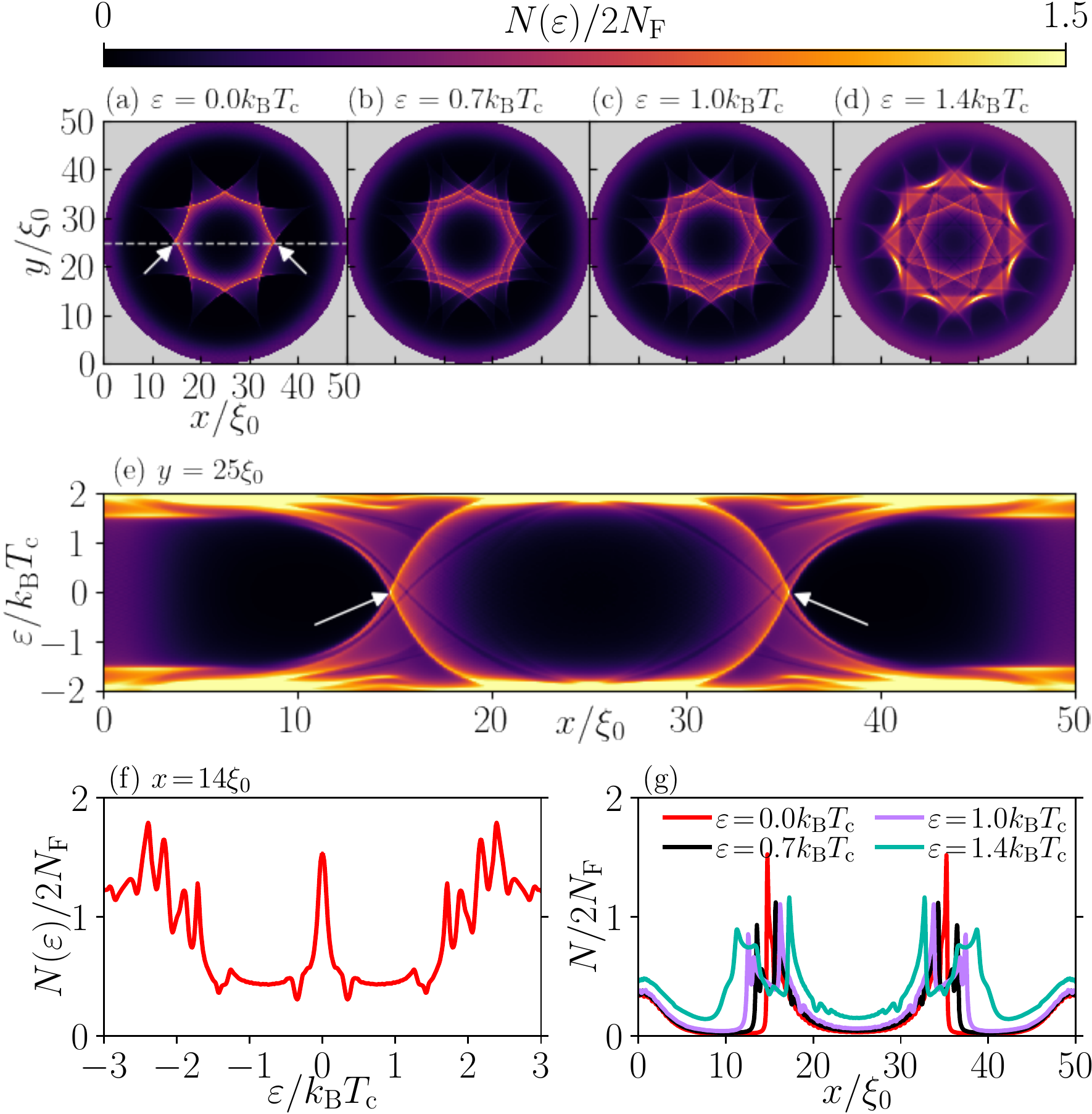}
	\caption{(a-d) Same as Fig.~2 in main text but for a CV with broken axial and rotational symmetries at $\Phiext=-8\Phi_0$ but otherwise the same parameters. (f) LDOS in the domain wall of CV (arrow in (a)). (g) LDOS across CV (dashed line in (a)) at fixed $\varepsilon$.
	}
	\label{fig:coreless_vortex:LDOS:ax_sym:break}
    \end{minipage}
\end{figure*}

In this SM section we provide complementary plots for CVs which spontaneously breaks axial and continuous rotational symmetries. They appear in a system where the dominant chirality $\Dplus$ ($\Dminus$) is exposed to negative (positive) external flux, as compared to the symmetry-preserving CV solutions for same dominant chiralities  exposed to positive (negative) external flux. In particular, in the main text we reported data for the symmetry-preserving CVs in both Figs.~1 and 2, while we summarize the most important results in Fig.~3 for the symmetry-broken CVs. For completeness we first provide the complete analogy for Figs.~1 and 2 for the symmetry-broken CVs. 
Thus, Fig.~\ref{fig:coreless_vortex:ax_sym:break} shows a CV with the same parameters and layout as in Fig.~1, but at $\Phiext=-8\Phi_0$. Visible here and not in the main text figures, is the nodal order parameter components which contain in total eight fractional vortices located in the domain wall.
Their locations are rotated with respect to in Figs.~1(a-d), due to an absence of the $\pi$-shift in the subdominant phase $\chi_-$, as seen when comparing also Figs.~\ref{fig:coreless_vortex:no_pi}(a-d). Moreover, we see that both the charge-current density and the flux density also display the octagon-shape patterns present in the order parameters.
However, at higher temperatures thermal broadening may make it  difficult to spot this broken symmetry in heatmaps of $|\Dplus|$, $|\Dminus|$, $\vj$, and $\vB$. 
In contrast, the broken symmetry and octagon-like shape is distinctly visible in the LDOS, as seen in Fig.~\ref{fig:coreless_vortex:LDOS:ax_sym:break} where we report the symmetry-broken CV LDOS with the same layout as in Fig.~2 for the symmetry-preserving CV and also show in (f,g) additional line cuts for an easier quantitative comparison of the LDOS peaks with Fig.~3 in the main text. For the latter we find a richer peak structure due to the broken symmetry. 
Finally, we also point out the absence of a zero-energy peak at the center of the CV, in contrast to the symmetric CV in Fig.~2. In the main text we stated that this peak only occurs in fully symmetric systems. We attribute it to hybridization of ballistic quasiparticle trajectories, crossing the domain wall of the CV. Technically speaking, the high symmetry of the perfect circular structure acts as a convex ``lense'' which focuses quasiparticle trajectories. Here we see that simply reducing the symmetry destroys this zero-energy peak, as it leads to concave domain walls which do not have the same lensing properties.

\begin{figure*}[hb]
	\includegraphics[width=\columnwidth]{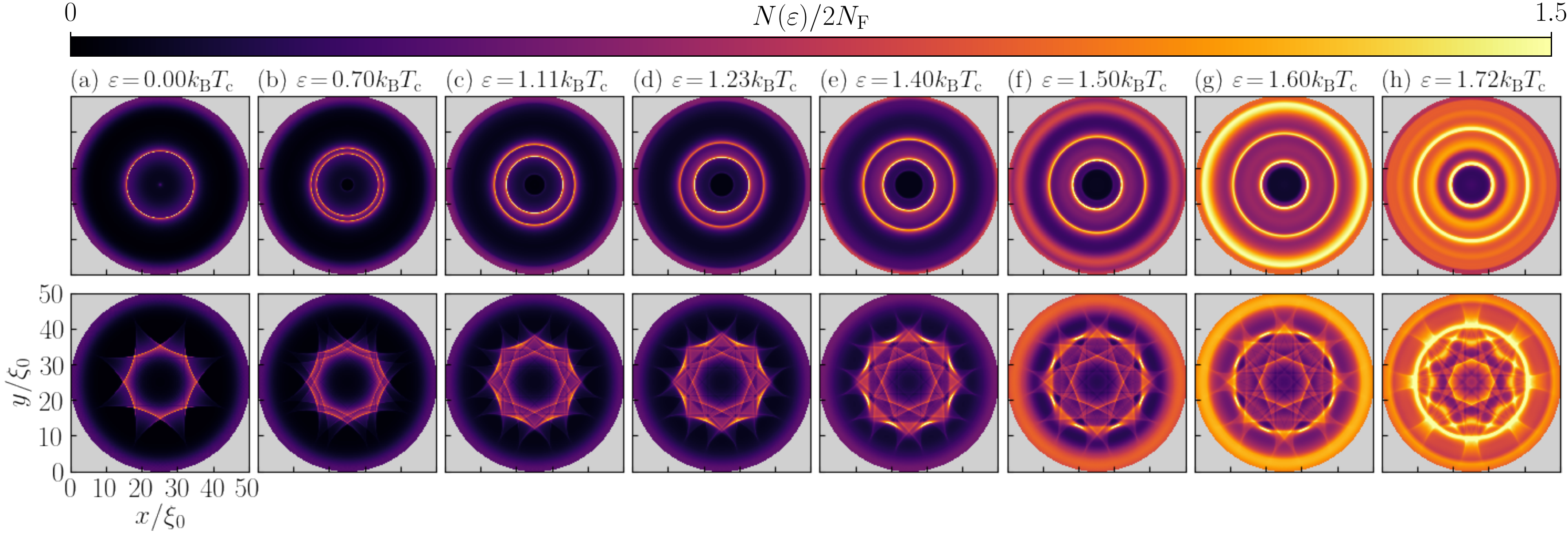}
	\caption{LDOS for a symmetric CV (top row) and symmetry-broken CV (bottom row). Same parameters as in Figs.~1,2 in the main text, but with $\Phiext=-8\Phi_0$ in the bottom row. }
	\label{fig:LDOS_extra_energies}
\end{figure*}

To further ease comparison between the symmetry-broken and symmetry-preserving CVs, we provide in Fig.~\ref{fig:LDOS_extra_energies} a side-by-side comparison between the LDOS of the symmetric CV (Fig.~2 in main text) and symmetry-broken CV (Fig.~4 in main text) for additional energies compared to the main text.
The figure illustrates that the strong differences between these two CVs established in the main text, remains throughout the whole sugbgap energy interval $\varepsilon \in [0,|\Delta_0|)$, where $|\Delta_0| \approx 1.76\kB\Tc$ is the maximum bulk gap. Specifically, the differences between continuous and discrete rotational symmetries (the latter set by the Chern number) is visible at all sub-gap energies. It also survives significant energy broadening as established in a later SM section.

\clearpage
\section{Energy broadening in realistic samples}
\label{app:broadening}
\begin{figure*}[b!]
    \centering
    \begin{minipage}[t]{0.49\textwidth}
	\includegraphics[width=\columnwidth]{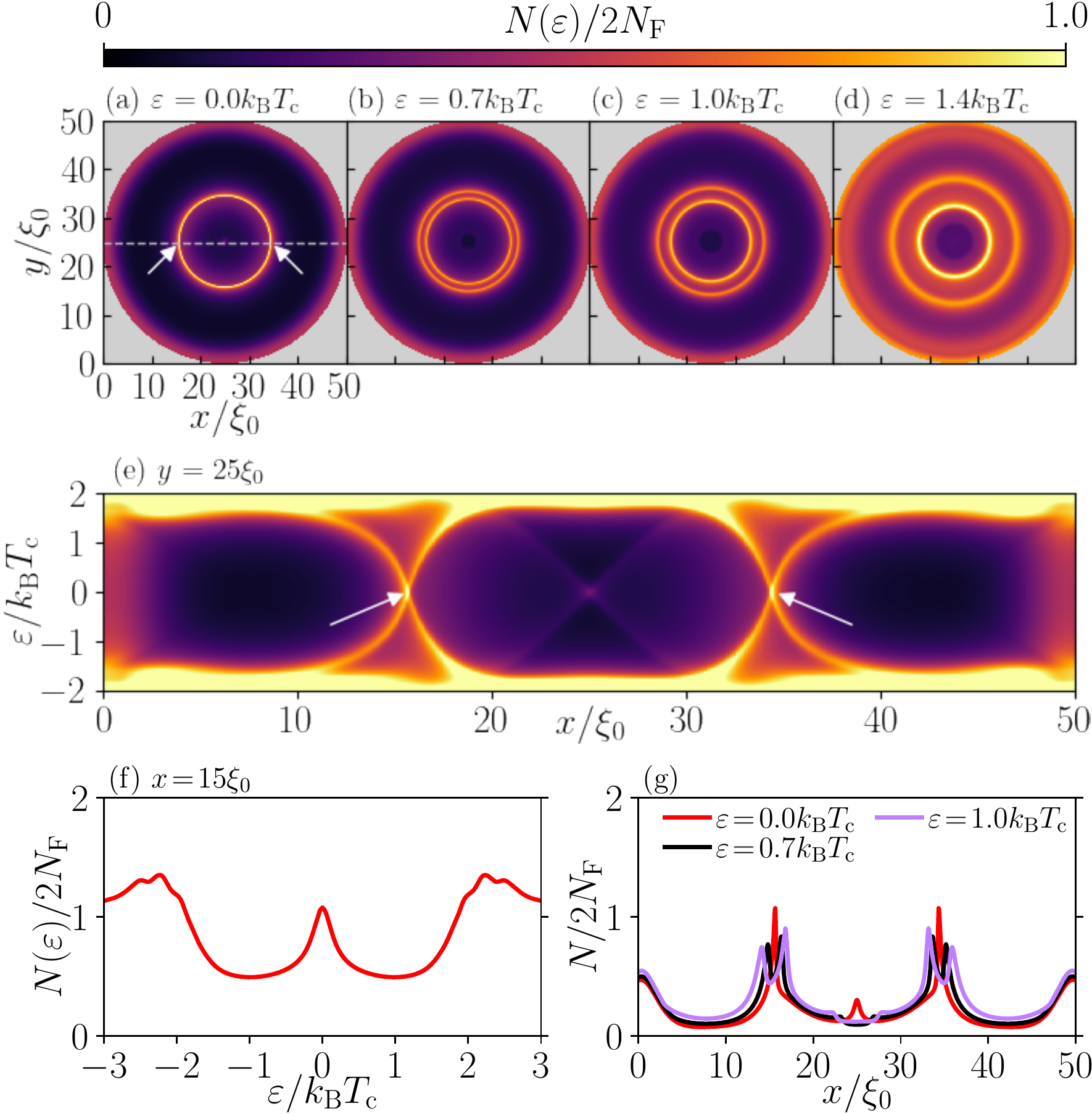}
	\caption{(a-d) Same as Fig.~2 in main text but with four times larger energy broadening $\delta=0.13\kB\Tc$.
	 (f) LDOS in the domain wall of CV (arrow in (a)). (g) LDOS across CV (dashed line in (a)) at fixed $\varepsilon$.}
	\label{fig:coreless_vortex:LDOS:broadening}
    \end{minipage}\hfill
    \begin{minipage}[t]{0.49\textwidth}
	\includegraphics[width=\columnwidth]{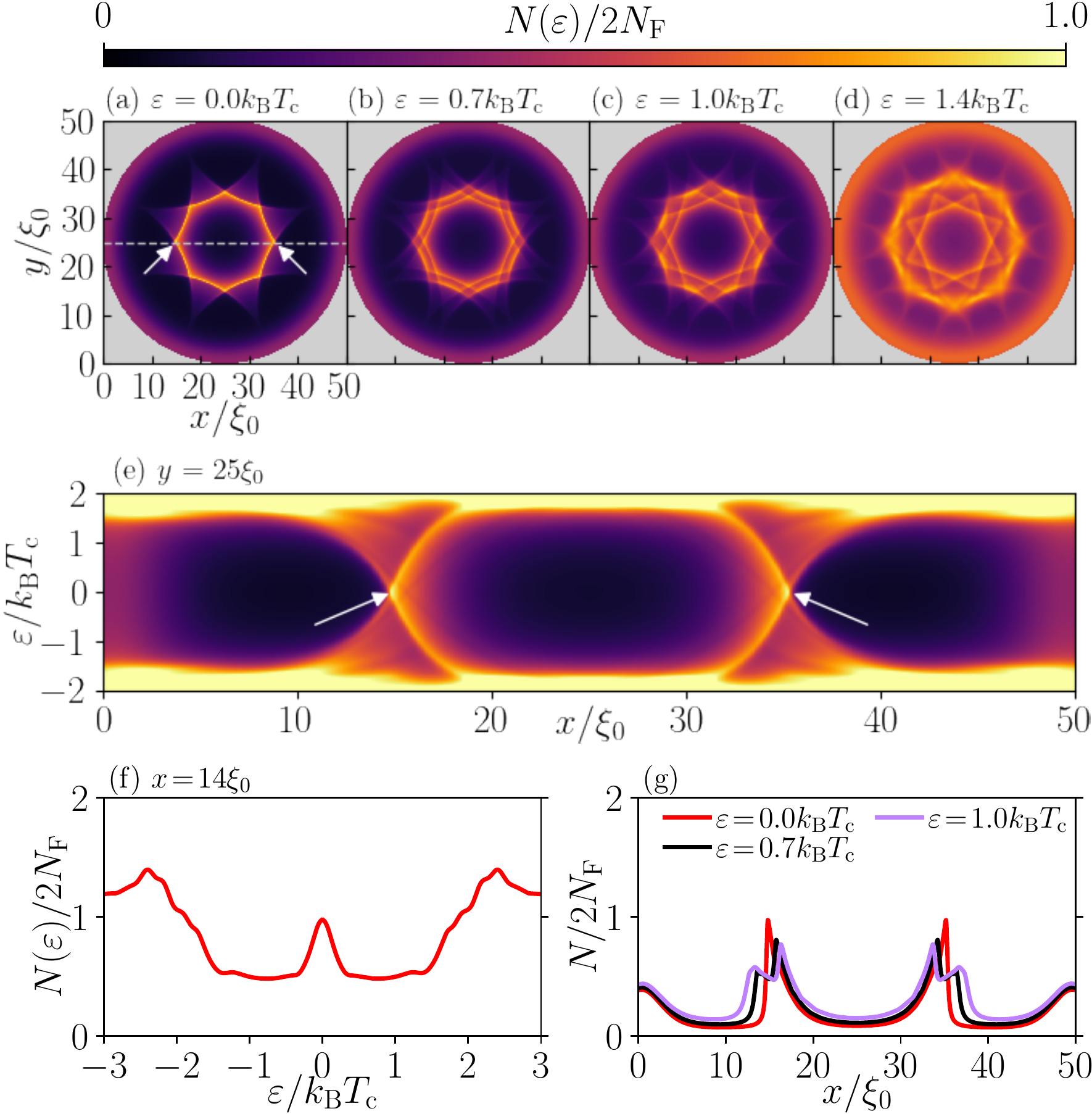}
	\caption{Same as Fig.~\ref{fig:coreless_vortex:LDOS:ax_sym:break} but with four times larger energy broadening $\delta=0.13\kB\Tc$.
	}
	\label{fig:coreless_vortex:LDOS:broadening:ax_sym:break}
    \end{minipage}
\end{figure*}

In the main text of this work we consider clean systems with perfectly specular boundaries, resulting in sharp peaks in the LDOS. The finite width and height of these peaks is only determined by an infinitesimal energy-broadening factor set to $\delta = 0.03\kB\Tc$, such that the LDOS $N(\epsilon)$ is evaluated at $\epsilon = \varepsilon + i\delta$, where $\varepsilon$ is the real-valued energy. It is however well-known that additional finite broadening can be caused by e.g.~disorder, impurity scattering, fluctuations, and interfaces with finite transparency \cite{Poenicke:1999,Lofwander:2001}. In order to investigate if our results are stable with respect to such broadening we in this SM section phenomenologically include a much larger broadening using $\delta \in [0.03, 0.31]\kB\Tc$,  where the upper limit is even comparable with the bulk superconducting gap $|\Delta_0|\approx1.76\kB\Tc$. In Figs.~\ref{fig:coreless_vortex:LDOS:broadening} and \ref{fig:coreless_vortex:LDOS:broadening:ax_sym:break} we reproduce Fig.~2 and Fig.~\ref{fig:coreless_vortex:LDOS:ax_sym:break} for the symmetry-preserving and symmetry-broken CVs, respectively, but now with $\delta=0.13\kB\Tc$. Clearly, the main features, such as the circular ring or octagon structures are still distinctly visible. We also for clarity include additional line cuts in (f) as a function of energy in the domain wall of the CV and in (g) across the CV at several different energies. These all show that the both the overall structure and peak heights remains distinguishable even for substantially increased broadening.
\begin{figure*}[b!]
    \centering
    \begin{minipage}[t]{0.49\textwidth}
	\includegraphics[width=\columnwidth]{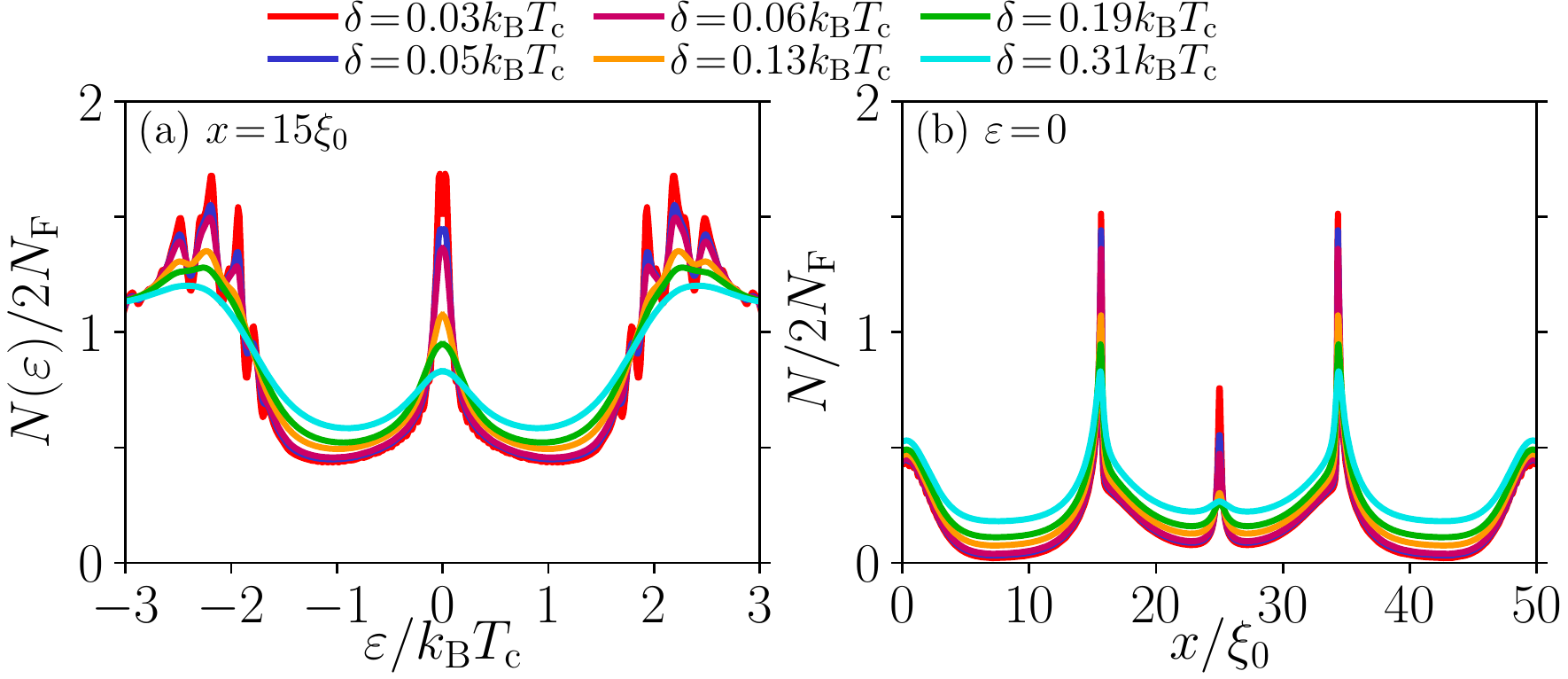}
	\caption{Line-cuts through LDOS plots in Fig.~\ref{fig:coreless_vortex:LDOS:broadening} at a point in the domain wall (left arrow in Fig.~\ref{fig:coreless_vortex:LDOS:broadening}) (a), and across the CV (dashed line in Fig.~\ref{fig:coreless_vortex:LDOS:broadening}) for zero energy (b), for a range of energy broadening $\delta$.
	}
	\label{fig:coreless_vortex:LDOS:broadening:cuts}
    \end{minipage}\hfill
    \begin{minipage}[t]{0.49\textwidth}
	\includegraphics[width=\columnwidth]{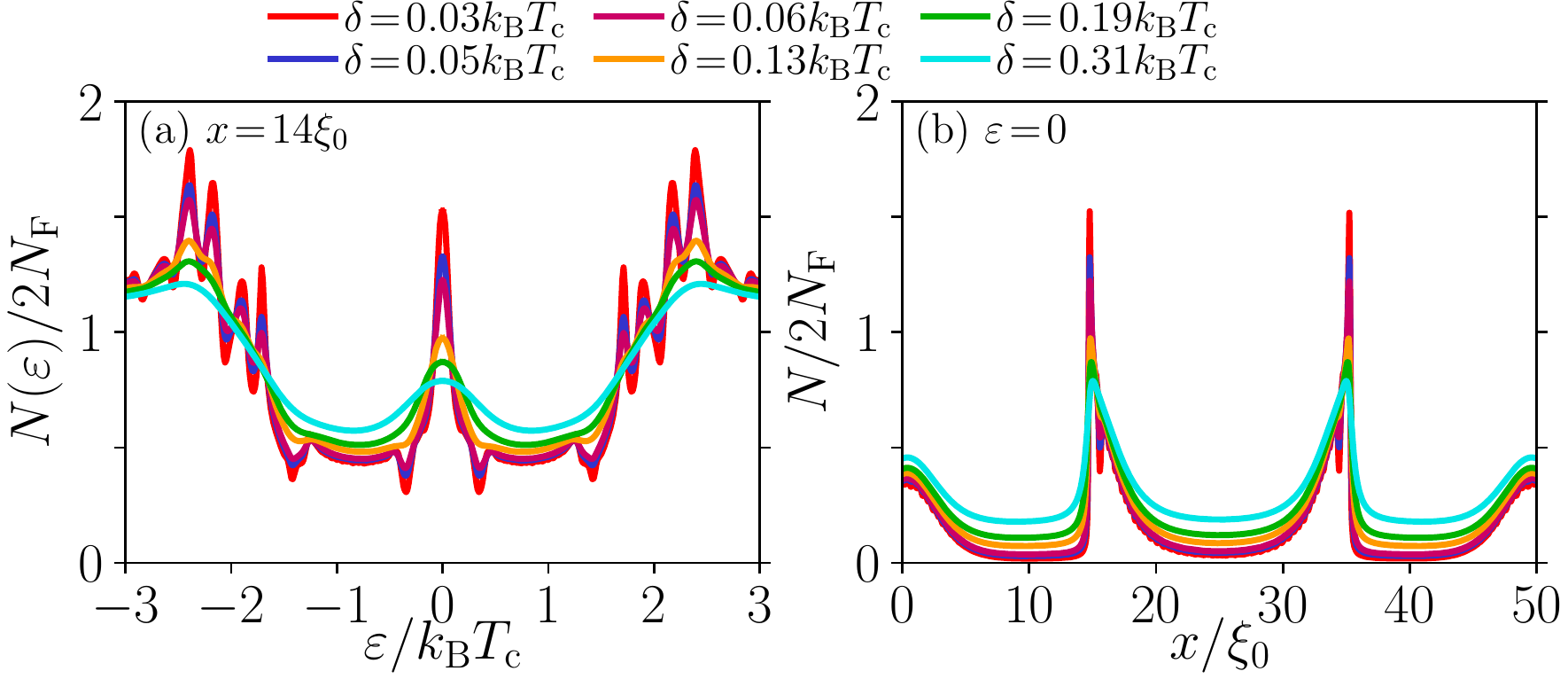}
	\caption{Line-cuts through LDOS plots in Fig.~\ref{fig:coreless_vortex:LDOS:broadening:ax_sym:break} at a point in the domain wall (left arrow in Fig.~\ref{fig:coreless_vortex:LDOS:broadening:ax_sym:break}) (a), and across the CV (dashed line in Fig.~\ref{fig:coreless_vortex:LDOS:broadening:ax_sym:break}) for zero energy (b), for a range of energy broadening $\delta$.
	}
	\label{fig:coreless_vortex:LDOS:broadening:ax_sym:break:cuts}
    \end{minipage}
\end{figure*} 

In Fig.~\ref{fig:coreless_vortex:LDOS:broadening:cuts} and \ref{fig:coreless_vortex:LDOS:broadening:ax_sym:break:cuts} we further explore the same LDOS peaks by plotting line-cuts in (a) in a point in the domain wall and in (b) at zero energy across the CV for the full range of broadening. We see that the domain wall peak structure remains clear even at $\delta=0.31\kB\Tc$, even though the peak is subdued. At the same time, we note that a larger flux (or temperature) leads to significantly larger LDOS peaks, which can thus be used to enhance signals.
Taken together, this leads us to conclude that the LDOS signatures of the CVs should survive in realistic scenarios where energy broadening of the spectrum is present due to various effects. \newtext{The companion article demonstrates the robustness further, where in particular energy broadening is exemplified by studying non-magnetic impurities \cite{Holmvall:2023:a}.}

\clearpage
\bibliographystyle{apsrev4-2}
\bibliography{supplemental.bib}